\documentclass[reprint, amsmath,amssymb,aps,twocolumn, prb,superscriptaddress]{revtex4-1}
\pdfoutput=1
\usepackage{amsfonts,amssymb}
\usepackage[subnum]{cases}
\usepackage{mathrsfs}
\usepackage{amsmath}
\usepackage[none]{hyphenat}
\usepackage{hyperref}    
\usepackage{bm}          
\usepackage{natbib}
\usepackage{soul} 
\usepackage{color}
\usepackage{longtable}
\usepackage{ textcomp }
\usepackage{verbatim}
\usepackage{fancyhdr} 
\usepackage{lastpage} 
\usepackage{extramarks} 
\usepackage{graphicx} 
\usepackage{lipsum} 
\usepackage{ amssymb }
\usepackage{slashed}
\usepackage{tikz}
\usepackage{comment}
\usepackage{natbib}
\usepackage[caption=false]{subfig}
\usepackage{ marvosym }
\usepackage{array}

\usepackage{amsthm} 

\newcommand{\abs}[1]{\left| #1 \right|} 
 
\let\baraccent=\= 
\renewcommand{\=}[1]{\stackrel{#1}{=}} 

\theoremstyle{definition}

\theoremstyle{remark}

\newcolumntype{C}[1]{>{\centering\let\newline\\\arraybackslash\hspace{0pt}}m{#1}}

\begin{document}
	\title{Anomalous Thermodynamic Properties of Electron Accumulation Layer in $\text{SrTiO}_3$}
	
	\author{Michael Sammon}
	\email{sammo017@umn.edu}
	\affiliation{Fine Theoretical Physics Institute, University of Minnesota, Minneapolis, MN 55455, USA}
	\author{Han Fu}
	\affiliation{Fine Theoretical Physics Institute, University of Minnesota, Minneapolis, MN 55455, USA}
	\author{B. I. Shklovskii}
	\affiliation{Fine Theoretical Physics Institute, University of Minnesota, Minneapolis, MN 55455, USA}
	\date{\today}
	\begin{abstract}
		Due to the nonlinear dielectric response within SrTiO$_{3}$ (STO), an accumulation layer created by positive charges at the surface of the STO sample ($x=0$) has an electron density profile $n(x)$ that slowly decays as $1/x^{12/7}$. Here we show that the long tail of $n(x)$ causes the magnetization and the specific heat of the accumulation layer to diverge at large $x$. We explore the truncation of the tail by the finite sample width $W$, the transition from the nonlinear to linear dielectric response with dielectric constant $\kappa$, and the use of a back gate with a negative voltage $-\abs{V}$. We find that both the magnetization and specific heat are anomalously large and obey nontrivial power law dependences on $W$, $\kappa$, or $\abs{V}$. We conclude with a discussion of how the capacitance as a function of the back gate voltage may be used to study the shape of the $n(x)$ tail in thin samples. 
	\end{abstract}
	\maketitle
	\section{Introduction}
	
	There is growing interest in the investigation of $\mathrm{ABO_3}$ perovskite crystals, which are important for numerous technological applications and show intriguing magnetic, superconducting, and multiferroic  properties \cite{Oxides_rev}.  Special attention \cite{Stemmer_STO,Zubko_oxides} is paid to heterostructures involving $\mathrm{SrTiO_3}$ (STO) which is a  semiconductor  with a band gap $E_g\simeq \mathrm{3.2~eV}$  \cite{Optical_absorbtion_STO} and a large dielectric constant $\kappa =2 \cdot 10^4$ at liquid helium temperatures. STO can be used as a building block for different types of devices, with reasonably large mobility \cite{Ohtomo_2004,Hwang_mobility}.
	
\begin{figure}
		\includegraphics[width=0.8\linewidth]{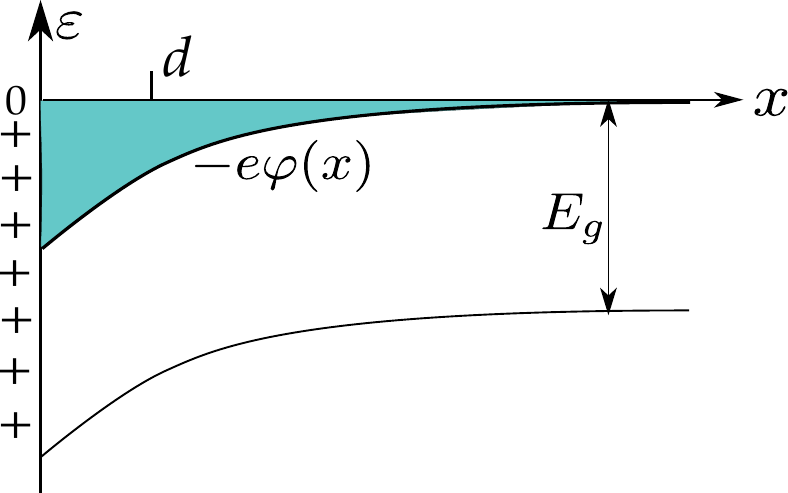}
		\caption{(Color  online) Schematic electron potential energy $-e\varphi(x)$ diagram of an accumulation layer in a moderately $n$-doped STO where $x$ is the distance from the surface. The electron (blue) is attracted by the positive charges (pluses) at $x=0$. The characteristic width of the gas is $d$. In the bulk of STO the Fermi level   $\varepsilon_F $ is near the bottom of the conduction band. Other positions of the Fermi level in the bulk are discussed in Sec. V.}
		\label{fig:accumulation}
	\end{figure}

	Many devices are based on the accumulation layer of electrons near a heterojunction interface in a moderately $n$-type doped STO. For example, one can get an accumulation layer with two-dimensional (2D) concentration $N=3\times 10^{14}$ cm$^{-2}$ of electrons on the STO side of the GTO/STO heterojunction induced by the electric field resulting from the ``polar catastrophe" in GdTiO$_3$ (GTO) \cite{Stemmer_GdTO} (see Fig. \ref{fig:accumulation}). The role of GTO can also be played by perovskites LaAlO$_3$ \cite{Ohtomo_2004,Hwang_mobility,Stemmer_STO}, NdAlO$_3$, LaVO$_3$ \cite{LaVO_STO}, SmTiO$_3$, PrAlO$_3$, NdGaO$_3$ \cite{different_polar_STO}, LaGaO$_3$ \cite{LaGaO_STO}, and LaTiO$_3$ \cite{LaTO_STO}. One can  accumulate an electron gas using a field effect \cite{10_percent,Hwang_gate,Stemmer_concentration_interface}. In Refs. \onlinecite{induced_superconductivity,Gallagher_2014} the authors accumulated up to $10^{14} ~\mathrm{cm}^{-2}$ electrons on the surface of STO using ionic liquid gating. Inside bulk STO $\delta$-doping by large concentrations of donors can be used  to introduce two accumulation layers of electrons  \cite{delta_doped_stemmer,delta_doped_STO_Hwang,delta_doped_STO_Stemmer}.
	Not surprisingly, the potential and electron density depth profiles in such devices have attracted a lot of attention \cite{Hwang_Xray,Hwang_PL,LAO_STO_Berreman,Stemmer_GdTO,MacDonald_theory,induced_superconductivity, abinitio_STO,abinitio_STO_2,distribution_LAO_STO,superconductivity_LAO_STO}.
	
	In order to describe the accumulation layer, we imagine that the effect of the doping, gate, or polar catastrophe can be thought of as a concentration $N$ of positive charge that lies at the STO surface. This charge attracts electrons to the surface, creating the accumulation layer illustrated in Fig. 1. In Ref. \onlinecite{RS}, authors calculated the three-dimensional (3D) electron density profile $n(x)$ of the accumulation layer with a large 2D density $N$ as a function of the distance $x$ from the surface. To account for the nonlinear dielectric response in STO they used the Landau-Ginzburg free energy expansion \cite{Ginzburg_ferroelectrics, Landau_stat} while describing the degenerate electron gas with the Thomas-Fermi approximation \cite{Thomas_Fermi}. They arrived at the self-consistent potential $\varphi(x)$
	\begin{equation}\label{eq:potential_nonlinear}
	\varphi(x)=\frac{C_1}{A^{2/7}}\frac{e}{b}\left(\frac{b}{a}\right)^{8/7}\left(\frac{b}{x+d}\right)^{8/7}
	\end{equation}
	and the electron density profile
	\begin{equation}
	n(x)=\frac{C_2}{A^{3/7}}\frac{1}{b^3}\left(\frac{b}{a}\right)^{12/7}\left(\frac{b}{x+d}\right)^{12/7}\label{eq:nonlinear_concentration},
	\end{equation}
	where $a=3.9$ $\AA $ is the lattice constant, $b=\hbar^2/m^*e^2= 0.30$ $\AA$, $m^*=1.8m_e$ is the effective mass of the electron, $m_e$ is the electron mass, and $d$ is the characteristic decay length of the electron density
	\begin{equation}
	d=\frac{C_3}{A^{3/5}}b\left(\frac{a}{b}\right)^{2/5}\left(Na^2\right)^{-7/5}\label{eq:decay_length}.
	\end{equation}
	Here $C_1$, $C_2$, $C_3$ are dimensionless constants, and $A$ is a numerical constant describing the nonlinear dielectric response. The values of these parameters and all other numerical constants can be found in Tab. \ref{tab:coefficients} in Appendix B.

\begin{figure}[h!]
	\begin{tabular}{c c}
		\subfloat{(a)\label{fig:no truncation}
			\includegraphics[width=0.4\linewidth]{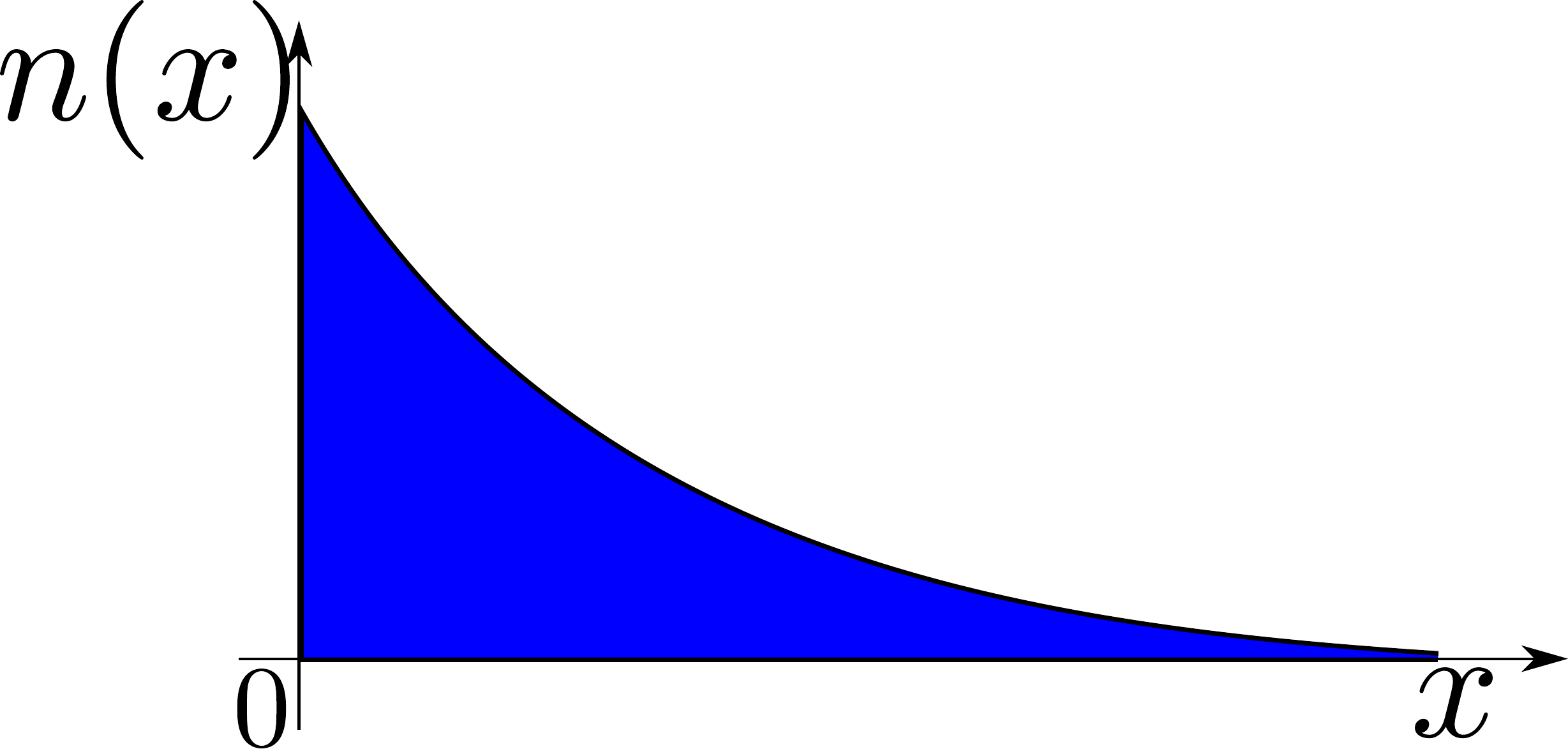}	
		}
		&	
		
		\subfloat{(b)\label{fig:finite width}
			\includegraphics[width=0.4\linewidth]{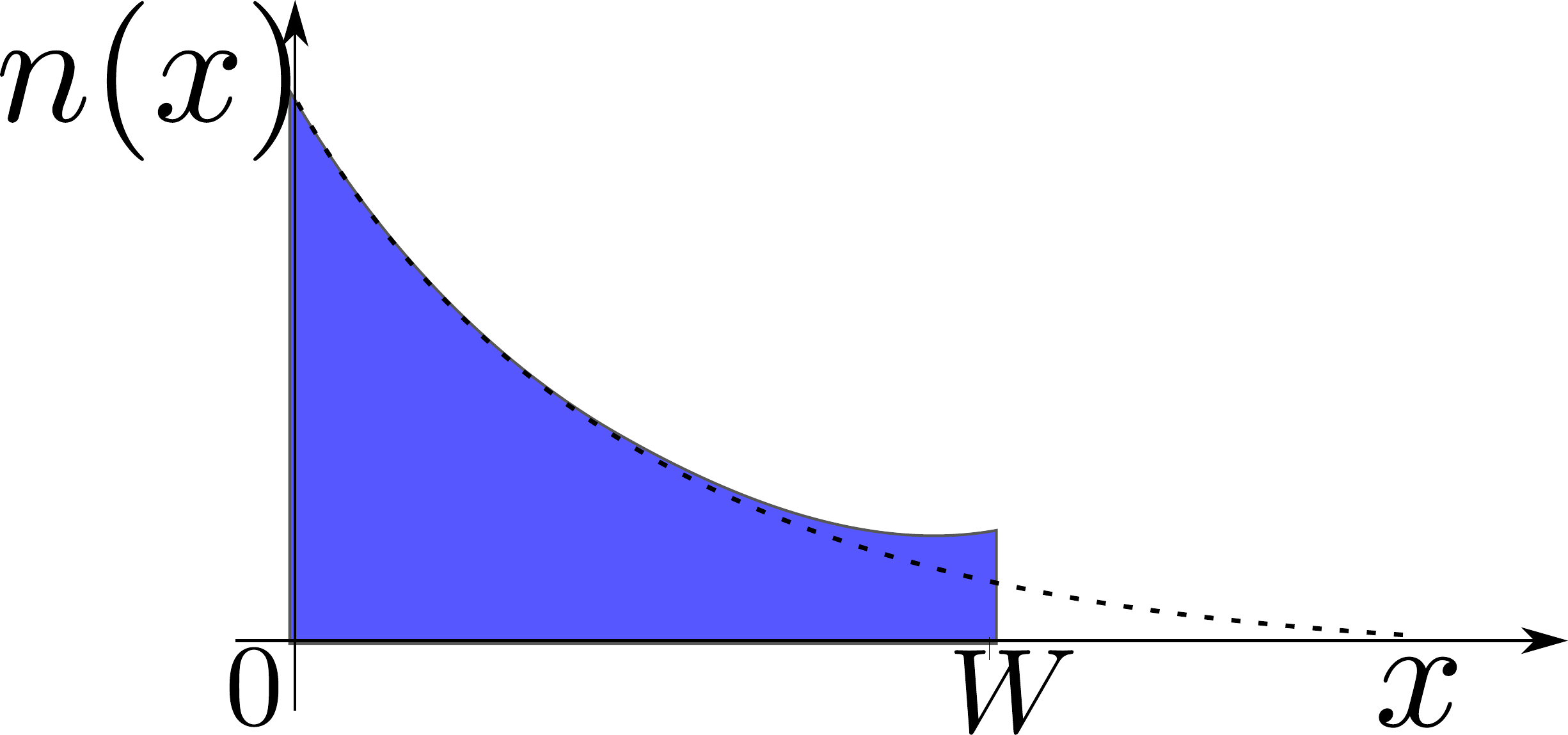}
		}
		\\
		\subfloat{(c)\label{fig:linear-nonlinear}
			\includegraphics[width=0.4\linewidth]{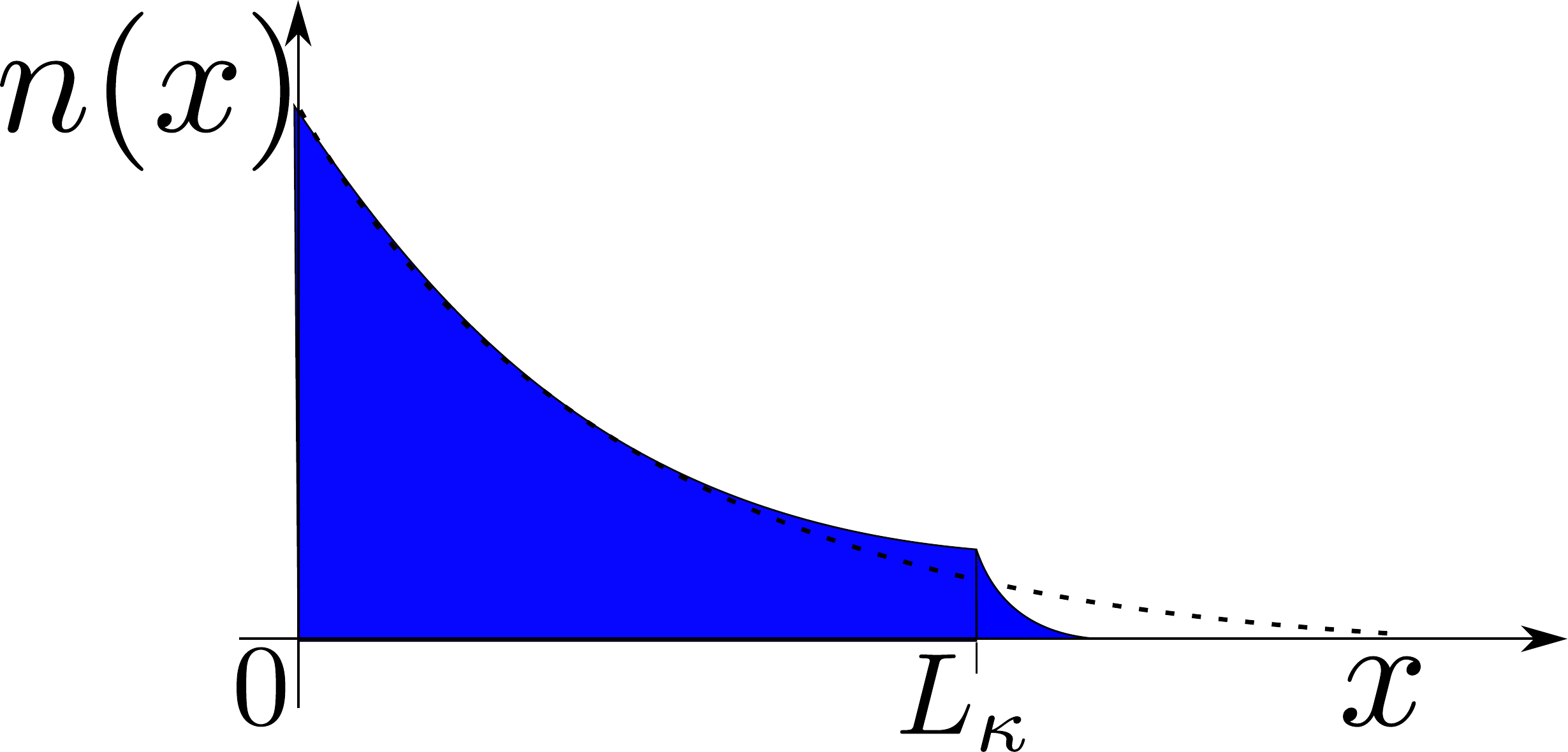}
		}
		&
		\subfloat{(d)\label{fig:back gate}
			\includegraphics[width=0.4\linewidth]{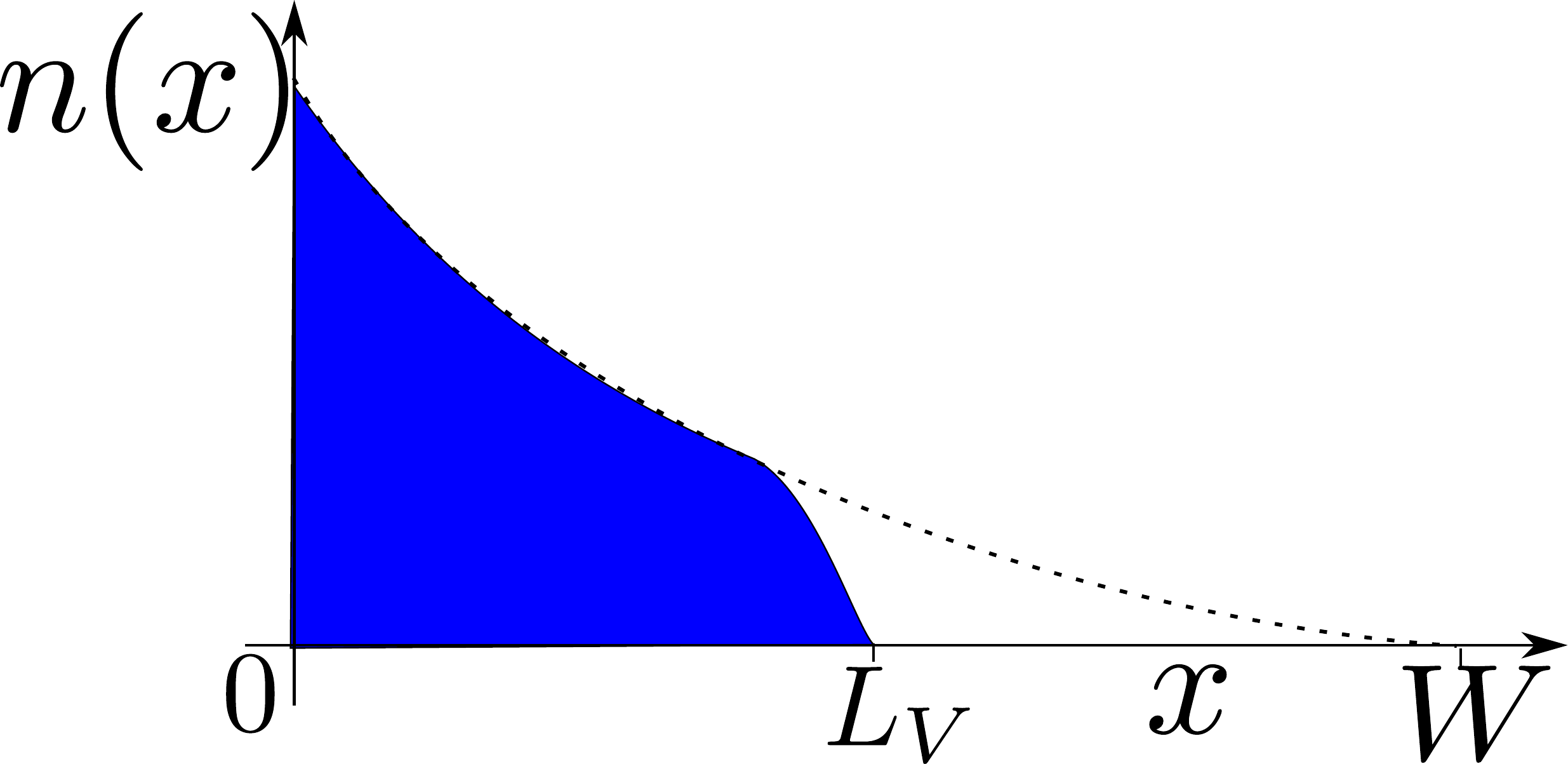}
		}
	\end{tabular}
	\caption{Schematics of the density profile $n(x)$ for an accumulation layer in STO with a) no truncation, b) truncation by the finite sample width $W$, c) truncation by the linear-nonlinear crossover, and d) truncation by the back gate voltage $V$. Here $x$ is the distance from the surface. The dotted line in b), c), and d), correspond to the density profile without truncation.}\label{fig:truncations}
\end{figure}

	In this paper, we first study the low-temperature magnetization $M_s$ and the specific heat $c_s$ per unit area of such an accumulation layer. Because the neutrality condition
	\begin{equation}
	N=\int_0^{\infty}  n(x) dx \label{eq:surface concentration}
	\end{equation}
converges, one might suspect that $M_s$ and $c_s$ are similar to that of a degenerate electron gas in a uniform layer, with a thickness $d$, surface concentration $N$, and a bulk density $n(N)=N/d$. For the purpose of comparison, we denote these quantities in the uniform layer as $\tilde{M_s}$ and $\tilde{c_s}$. Instead, we find that $M_s$ and $c_s$ are strongly enhanced above $\tilde{M_s}$ and $\tilde{c_s}$. The reason for this is that in calculating these quantities, we must integrate the local magnetization $M(x)$ and specific heat $c(x)$ per unit volume across the entire layer. Both these quantities are proportional to the local density of states at the Fermi level, which decreases slowly as $n(x)^{1/3}\propto 1/x^{4/7}$. As a result, integrating $M(x)$ and $c(x)$ over the accumulation layer causes $M_s$ and $c_s$ to diverge, and the integral must be truncated at a large $x=L$. There are several possible mechanisms for the truncation, such as the finite width of the sample $W$, the crossover to a linear dielectric response with a dielectric constant $\kappa$, and the application of a back gate with negative voltage $-\abs{V}$ as shown in Fig. \ref{fig:truncations}. As a result, the magnetization $M_s(L)$ and the specific heat $c_s(L)$ per unit area of the accumulation layer are enhanced above their uniform layer values $\tilde{M_s}$ and $\tilde{c_s}$ by a power law factor that depends on the truncation length $L$. This introduces a power law dependence on the width $W$, the linear dielectric constant $\kappa$, and the magnitude of the back gate voltage $\abs{V}$ depending on which mechanism is responsible for the truncation. Similar anomalous behavior of kinetic coefficients for STO accumulation layers dominated by surface scattering has previously been studied.\cite{RAT} We emphasize that in this paper we are only discussing thermodynamic properties of the electrons in which the different scattering mechanisms play no role. 

The second half of the paper describes the capacitance formed between the accumulation layer and a back gate located at the $x=W$ edge of the sample. When the magnitude of the voltage $\abs{V}$ is small, the capacitance can be described with the usual Debye screening radius and an effective dielectric constant determined by the electric field $E(W)$ at the sample edge. We find that the capacitance in this region is approximately constant with respect to the back gate voltage. However, when the voltage is increased beyond the limit of the Debye theory, the electrons are confined to a region of thickness $L_V$ as measured from the $x=0$ surface, leaving a fully depleted region of size $W-L_V$ near the back gate. We show that the capacitance in this limit can be described with an effective width $W+\beta L_V$ and a dielectric constant $\kappa(V)\propto V^{-2/3}$ that changes with the back gate voltage. What is surprising is that the coefficient $\beta$ is positive, leading to an effective width larger than the width of the sample. As we explain in detail below, this counterintuitive result comes from a combination of the dependence of $L_V$ on $V$ and the dependence of the dielectric constant on $L_V$.

The paper is organized as follows. In Sec. II we describe the magnetization of the gas in both the uniform layer and the accumulation layer, and describe the different truncation mechanisms. In Sec. III we repeat the same discussion for the specific heat and summarize the results in Table \ref{table:specific heat}. In Sec. IV we discuss how capacitance measurements as a function of the negative back gate voltage $-\abs{V}$ may be used to measure the truncation length $L_V$ and the concentration $n(W)$ and thus as an experimental verification of the tail of the distribution. Finally we conclude in Sec. V. 

    \section{Magnetization}

	Let us explore the magnetization of an STO accumulation layer.  We assume that the magnetic field is applied perpendicular to the surface, and is weak in the sense that $\mu_{B} B\ll k_{B}T$, where $\mu_{B}=\abs{e}\hbar/2m_ec$ is the Bohr magneton, $T$ is the temperature, and $k_B$ is the Boltzmann constant. We know that under these conditions there are two contributions to the magnetization of a degenerate electron gas; the paramagnetic effect from the spins of the electrons and the diamagnetic effect due to the orbital motion of electrons in the applied magnetic field.

	First let us discuss Pauli paramagnetism within the accumulation layer.\cite{Landau_stat}  In the weak field limit ($\mu_{B}B\ll E_{F}$) the magnetization per unit volume is given by 
	\begin{equation}
	M=\mu_{B}^2Bg(n) \label{eq:Volume Magnetization}
	\end{equation}
	where 
	 \begin{equation}\label{eq:density of states}
	g(n)=\frac{m^*}{\pi^2\hbar^2}(3\pi^2n)^{1/3}=\frac{3}{2}\frac{n}{E_F(n)}
	\end{equation}
	 is the density of states at the Fermi level and we have used $E_F=(\hbar^2/2m^*)(3\pi^2n)^{2/3}$ to relate the density to the Fermi energy. This formula has a very simple interpretation. The Zeeman splitting of the different spins gives rise to an additional occupancy of electrons whose spin is aligned with the magnetic field. In the weak field limit, the response is linear, so that each electron within an energy range $\mu_B B$ of the Fermi level contributes a moment $\mu_B$ to the magnetization. The total density of electrons that contribute is then $\mu_{B} Bg(n)$, giving rise to Eq. (\ref{eq:Volume Magnetization}). It is important to note that Eq. (\ref{eq:Volume Magnetization}) is valid both in the uniform layer and the accumulation layer, so long as we identify $n=n(x)$ as the bulk density at a distance $x$ from the surface, and $M=M(x)$ as the magnetization per unit volume at the distance $x$.

	 In order to calculate the total magnetization per unit area $M_s$,  $M(x)$ must be integrated over the entire layer:
	\begin{equation}
	M_s=\int_0^{\infty}  M(x)dx. \label{eq:layer magnetization}
	\end{equation}
	 Using this definition, we now discuss how $M_s$ differs between the uniform layer and the accumulation layer.

	In the case of a uniform layer of thickness $d$ and bulk density $n(N)=N/d$, Eq. (\ref{eq:layer magnetization}) shows that the magnetization is 
	\begin{equation}
	\tilde{M_s}=\alpha\frac{3}{2}\mu_B^2 B\frac{N}{E_F(N/d)} , \label{eq:uniform layer}
	\end{equation}
	where $E_F(N/d)$ is the Fermi energy of a uniform layer with bulk density $N/d$. Additionally, we have introduced a correction factor $\alpha<1$ to account for the diamagnetic contribution(see below).

	When the layer is not uniform, we use Eq. (\ref{eq:nonlinear_concentration}) for the local density. The magnetization of the accumulation layer is then found to be
	\begin{equation}
	M_s\propto\int_0^{\infty}\frac{1}{(x+d)^{4/7}}dx.
	\end{equation}
	We see that the integral diverges as $x^{3/7}$ for large $x$, and so we truncate the integral at a value $x=L$. With this truncation the leading order contribution to the magnetization per unit area is found to be
	\begin{gather}
	M_s(L)=C_4  \tilde{M_s} \left(\frac{L}{d}\right)^{3/7}, \label{eq:surface Magnetization}
	\end{gather}
	where $\tilde{M_s}$ is defined by Eq. (\ref{eq:uniform layer}) and $C_4$ is a numerical constant. 
	
	The truncation length $L$ can be a result of i) the finite width of the sample, ii) truncation due to the transition to a linear dielectric response, and iii) the application of a back gate to the layer. The details of each truncation mechanism will be discussed individually below and the smallest of these values is to be substituted into Eq. (\ref{eq:surface Magnetization}).

	\textit{Finite Sample Width}. In a sample with a very small width, such as GTO/STO/LSAT heterostructures with an STO layer of width $W$, the truncation is due to the finite sample width. Here LSAT stands for (LaAlO$_3$)$_{0.3}$(Sr$_2$AlTaO$_6$)$_{0.7}$ In this case the magnetization is given simply by 
	\begin{equation}
	M_s(W)=C_4\tilde{M_s}\left(\frac{W}{d}\right)^{3/7}. \label{eq:Magnetization Finite Width}
	\end{equation}
	
	For symmetric quantum wells such as GTO/STO/GTO with an STO layer of width $W$, an accumulation layer forms on each GTO/STO interface and the density profile is symmetric about the center of the well\cite{HanSTO}. When $W>8a=3.2$ nm, these accumulation layers are essentially separate and one can calculate the magnetization using the above method for each of the layers, using a truncation length of $W/2$ instead of $W$.

	\textit{Transition to linear dielectric response}. Within the layer, the electric field decays with increasing $x$ as $1/x^{15/7}$. As a result the field at large $x$ becomes so small that the dielectric response of the STO sample becomes linear with a large dielectric constant $\kappa$. It has been shown in Ref. \onlinecite{RAT} that this crossover occurs at a distance
	\begin{equation}
	L_{\kappa}=C_5b\kappa^{7/10}\left(\frac{a}{b}\right)^{2/5}\label{eq:linear truncation}
	\end{equation}
	where $C_5$ is a numerical constant.\cite{RAT} Substituting this into Eq. (\ref{eq:surface Magnetization}), the magnetization of the layer becomes
	\begin{equation}
	M_s(\kappa)=C_6\tilde{M_s}\left(\frac{b}{d}\right)^{3/7}\left(\frac{a}{b}\right)^{6/35}\kappa^{3/10}\label{eq:Magnetization Linear Cutoff}
	\end{equation}
	where $C_6$ is a numerical constant.
	
	\textit{Truncation by the back gate voltage}. In an STO sample of width $W$, a back gate can be used to apply a voltage $V$ to the gas and alter the structure of the layer. When $V<0$ electrons are repelled away from the back gate\footnote{When $V>0$ extra electrons are brought to STO. They form an accumulation layer at the interface with a gate, similar to the symmetric wells GTO/STO/GTO As a result the magnetization and specific heat will grow with $V$ and in the symmetric case are enhanced by a factor $2^{4/7}$ compared to when $V=0$ and the truncation was by the finite sample width $W$.}. Let us assume that $n(x)$ vanishes at $x=L_V$, and that $L_V\ll W$, where $W$ is the width of the STO sample. Then we can think that the magnitude of the back gate electric field is $E_x=\abs{V}/W$.  Conversely, we mentioned before that the electric field within the accumulation layer $E_x=-d\varphi/dx$ decays like $1/x^{15/7}$, where $\varphi(x)$ is given by Eq. (\ref{eq:potential_nonlinear}) . The length $L_V$ can then be defined as the distance in which these two electric fields are equal and we find
	\begin{equation}
	L_{V}=\gamma b\left(\frac{b}{a}\right)^{8/15}\left(\frac{\abs{V}b^2}{eW}\right)^{-7/15}.\label{eq:Back Gate Truncation}
	\end{equation} 
	 Here we have introduced a numerical constant $\gamma$ which cannot be determined from the qualitative description above. A numerical calculation using the Thomas-Fermi approach finds $\gamma\approx 3.94$, and the details of the procedure are described in Appendix. 
	Using $L_V$ as the truncation length in Eq. (\ref{eq:surface Magnetization}), we arrive at the magnetization as a function of back gate voltage
	\begin{equation}
	M_s(V)=C_7\tilde{M_s}\left(\frac{b}{d}\right)^{3/7}\left(\frac{b}{a}\right)^{8/35}\left(\frac{eW}{b^2\abs{V}}\right)^{1/5} \label{eq:Magnetization Back Gate}
	\end{equation}
	where $C_7$ is a numerical constant.
	
	\textit{Diamagnetism}. Now we address the correction factor $\alpha$ due to the diamagnetic effect. In a uniform system, this effect leads to the well known value $-(1/3)\tilde{M_s}$, with one major difference. Because the diamagnetic effect is a result of the orbital motion, we must use the effective mass $m^{*}$ instead of the bare electron mass $m_e$ in the definition of the magnetic moment $\mu_{B}=\abs{e}\hbar/2m^*c$. Because $\tilde{M_s}\propto\mu_B^2$, we find that the correction factor $\alpha$ is then given by 
	\begin{equation}
	\alpha=1-\frac{1}{3}\left(\frac{m_e}{m^*}\right)^{2}.
	\end{equation}
	 In the case of STO, we use the fact that $m^*=1.8$ $m_e$ and find that $\alpha\approx0.90$.
	\section{Specific Heat}
	The specific heat per unit volume of a uniform gas at low temperatures is known to depend linearly on the temperature and described by the formula
	\begin{equation}
	c=\frac{\pi^2}{3}k_B^2T g(E_F). \label{eq:Uniform Specific Heat}
	\end{equation}
	 This equation is similar in nature and interpretation to that of the magnetization for a uniform gas, with $\mu_B B$ replaced by $k_BT$. 
	
		\begin{table}[t!]
			\begin{tabular}{|C{2.5cm}|c|}
				\hline
				 &\\[-5pt]
				Truncation   & $c_s(L)$ \\
				 &\\[-5pt]
				\hline \hline 
				 &\\[-5pt]
				
				Finite Sample Width W & $\displaystyle C_4\tilde{c_s}\left(\frac{W}{d}\right)^{3/7}$\\
				 &\\[-5pt]
				\hline
				&\\[-5pt]
				Crossover to Linear Dielectric Constant $\kappa$ &  $\displaystyle C_6\tilde{c_s}\left(\frac{b}{d}\right)^{3/7}\left(\frac{a}{b}\right)^{6/35}\kappa^{3/10}$\\
				&\\[-5pt]
				\hline 
				&\\[-5pt]
				Back gate Voltage $V$ & $\displaystyle C_7\tilde{c_s}\left(\frac{b}{d}\right)^{3/7}\left(\frac{b}{a}\right)^{8/35}\left(\frac{eW}{b^2\abs{V}}\right)^{1/5}$\\[10pt]
				\hline	
			\end{tabular}
			\caption{Specific heat per unit area $c_s$ of the STO accumulation layer for the different truncation mechanisms. Here $\tilde{c_s}$ is the specific heat per unit area of a degenerate gas in a uniform layer of thickness $d$ and bulk concentration $N/d$, $W$ is the width of the STO sample, $\kappa$ is the linear dielectric constant of STO, and $V$ is the back gate voltage. $C_4$, $C_6$, and $C_7$ are numerical constants.} 
			\label{table:specific heat}
		\end{table}
		
	Just as before, in order to describe the specific heat of the accumulation layer, Eq. (\ref{eq:Uniform Specific Heat}) must be expressed through its local value and integrated over the entire layer. Because the only dependence on position enters through the density of states, the divergence of the integration is identical to that of the magnetization. Therefore, it can easily be shown that the specific heat of the accumulation layer is given by 
	\begin{equation}
	c_s(L)=C_4\tilde{c_s}\left(\frac{L}{d}\right)^{3/7},
	\end{equation}
	where $\tilde{c_s}=(\pi^2/2)k_B^2TN/E_F(N/d)$ is the specific heat per unit area of the uniform layer of thickness $d$ and bulk density $N/d$, $C_4$ is the same numerical constant that appears in Eq. (\ref{eq:surface Magnetization}), and $L$ is the truncation length. The truncation mechanisms discussed in the previous section are the same for the specific heat leading to the results in Table \ref{table:specific heat}.

	\section{Back Gate Capacitance of Thin STO Samples}
	In the previous sections we have discussed the effect of the long tail of $n(x)$ on various thermodynamic quantities, and found that the magnetization and specific heat are enhanced by a factor proportional to $L^{3/7}$. In principle the dependence of $M_s$ and $c_s$ on the truncation length can be used for an experimental study of the tail of the distribution. Here we would like to describe how the capacitance as a function of the back gate voltage may also be used to study the tail of the distribution. A similar study of the quantum capacitance of the accumulation layer has previously been suggested as a tool to measure the characteristic length $d$ of the layer.\cite{RS} Earlier negative compressibility in LAO/STO structures was discovered by capacitance studies\cite{licapacitance,tinkl}.

	\begin{figure}

		\subfloat{\includegraphics[width=0.8\linewidth]{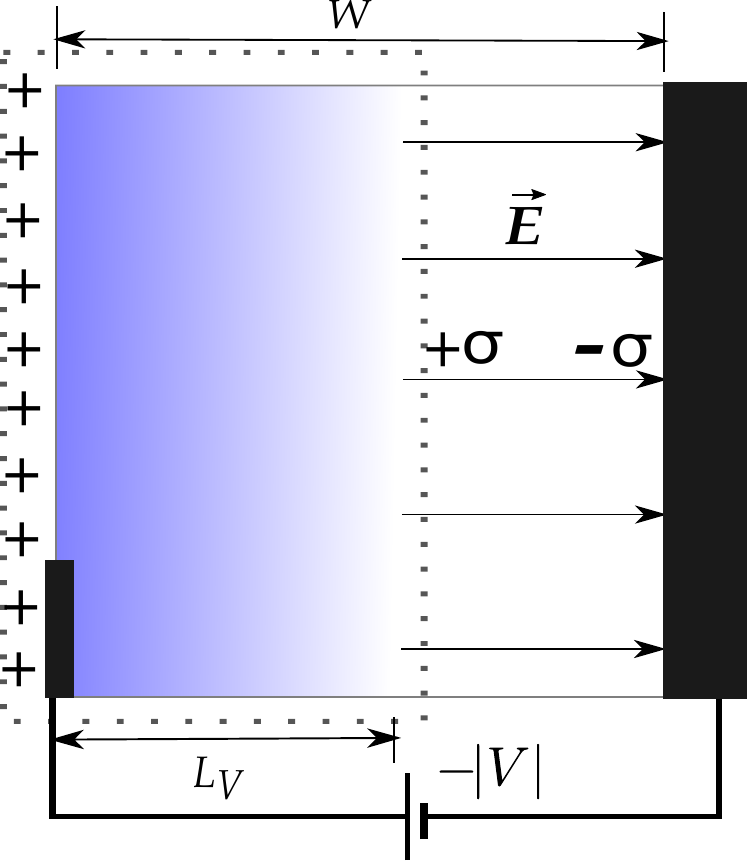}}
		
		\caption{Schematic image of the STO layer of width $W$ with a back gate at a negative voltage (right side). The back gate depletes the accumulation layer, so that the electron gas shown in blue (grey) lies in a region of size $L_V$ and a depletion layer of size $W-L_V$ is created. A small number of electrons are "stolen" by the back gate resulting in a negative surface charge $-\sigma$, while the left side (dotted box) gains a net positive surface charge $\sigma$ and forms a capacitor with the back gate.} \label{fig:back gate schematic}
	\end{figure}

	In the following dicussion we assume that prior to the application of the back gate the tail is truncated by the sample width $W$. Let us now imagine that a negative back gate voltage $-\abs{V}$ is applied to the STO sample by a metallic gate mounted on the $x=W$ edge. Let us further assume that part of the tail has been depleted so that the length of the layer is $L_V\ll W$, where $L_V$ is defined in Eq. (\ref{eq:Back Gate Truncation}). This means that the back gate has ``stolen" a small amount of electrons and acquires a negative charge $-\sigma$ while leaving a net positive charge $\sigma=e(N-\int_0^{L_V}n(x)dx)$ on the STO side as illustrated in Fig. \ref{fig:back gate schematic}. Thus a capacitor is formed between the accumulation layer and the back gate with charge $\sigma$, voltage $V$, and the inverse differential capacitance per unit area $C^{-1}=dV/d\sigma$. 
	
	In order to calculate the differential capacitance $C^{-1}$, we must first relate the potential $V$ across the capacitor to the charge per unit area $\sigma$. We proceed as follows. The region between the plates is fully depleted of electrons. As a result, the electric field $E$ within this region is given by $E=\abs{V}/(W-L_{V})$. Additionally, we know from Gauss's law that the displacement field $D$ is such that $D=4\pi\sigma$. Using the Landau-Ginzburg description of the dielectric response of the STO lattice, one finds that $E=AD^3/((4\pi)^3P_0^2)$ where $P_0=e/a^2$ is the characteristic polarization of STO. Combining these three equations we find that
	\begin{equation}
	A\frac{\sigma^3}{(e/a^2)^2}=\frac{\abs{V}}{W-L_{V}}.\label{eq:nonlinear electric field}
	\end{equation}

		 Taking the derivative $dV/d\sigma$, we find
		\begin{equation}\label{eq:general capacitance}
		C^{-1}\simeq3A\frac{\sigma^2}{(e/a^2)^2}(W-L_V)-A\frac{dL_V}{d\sigma}\frac{\sigma^3}{(e/a^2)^2}.
		\end{equation}

		Using Eq. (\ref{eq:nonlinear electric field}) with Eq. (\ref{eq:general capacitance}) and keeping only the first order in $L_V/W$, we find the capacitance to be 
		\begin{equation}
		C^{-1}(V)=\frac{4\pi}{\kappa(V)}(W+\beta L_{V}),\label{eq:back gate capacitance}
		\end{equation}
		where $\beta=2/15$ and
		\begin{equation}
		\kappa(V)=\frac{4\pi}{3}\left(\frac{A^{1/2}\abs{V}a^2}{We}\right)^{-2/3}\label{eq:back gate dielectric}
		\end{equation}
		is the dielectric constant when the accumulation layer is fully depleted. 
		
		We see from Eq. (\ref{eq:back gate capacitance}) that the correction to first order in $L_{V}$ increases the effective width. The fact that this correction is positive may seem counterintuitive as the effective thickness of the capacitor becomes larger than the sample width $W$. However, we see from Eq. (\ref{eq:general capacitance}) that this is not the real width of the capacitor. Instead the positive correction comes from the combined dependence of $L_V$ on the charge $\sigma$ and the dependence of the dielectric constant on $L_V$ to give an overall positive correction to the main term of order $W$.  If one wishes to verify Eq. (\ref{eq:back gate capacitance}) and the positive correction, one may plot $[4\pi C_{V}/\kappa(V)]^{-1}-W$ vs. $V$ and examine whether it agrees with the $\abs{V}^{-7/15}$ behavior given by Eq. (\ref{eq:Back Gate Truncation}).

	If the applied voltage is sufficiently small so that $L_V=W$ then there is no fully depleted region. Instead we can linearize the small depletion of the layer around the density $n(W)$ at the right edge of the sample. The capacitance is then given by the familiar expression  
	\begin{equation}
	C^{-1}=\frac{4\pi R_D}{\kappa_{eff}}
	\end{equation}
	where 
	\begin{equation}
	R_D^2=\frac{\kappa_{eff}}{4\pi e^2g(W)}
	\end{equation}
	is the Debye screening radius, $g(W)$ is the density of states at $x=W$, and $\kappa_{eff}=(1/3)D/E$ is the effective dielectric constant defined by the derivative $\delta D/\delta E$ at the right edge. Using $E=AD^3/((4\pi)^3P_0^2)$, $E=-d\varphi/dx$ with $\varphi(x)$ from Eq. (\ref{eq:potential_nonlinear}), we find 
	\begin{equation}\label{eq:dielectric debye}
	\kappa_{eff}=\frac{C_{\kappa}}{A^{1/7}}\left(\frac{b}{a}\right)^{4/7}\left(\frac{W}{b}\right)^{10/7}
	\end{equation}
	where $C_{\kappa}$ is a numerical coefficent. Combining Eq. (\ref{eq:dielectric debye}) with $g(W)=g(n(W))$ from Eq. (\ref{eq:density of states}), we find that 
	\begin{equation}\label{eq:capacitance debye}
	C^{-1}=C_8\frac{4\pi W}{\kappa_{eff}}
	\end{equation}
	where $C_8$ is a numerical coefficient. If we compare Eq. (\ref{eq:capacitance debye}) with our earlier expression Eq. (\ref{eq:back gate capacitance}), we see that as the magnitude of the voltage is decreased, the capacitance first grows, and then saturates at a constant value related to the electron density $n(x)$ near the sample edge. Therefore measurements of the peak capacitance near zero voltage allow for a study of the tail of the density distribution.  
	
	Let us discuss in more detail the necessary conditions for this to be observed. In the above discussion, we assumed that the truncation prior to the application of the back gate was by the sample width $W$. This need not be the case, as when the sample width becomes too large the main truncation will be due to the crossover to a linear dielectric response. This does not change any of the results, so long as we require that $L_{V}\ll L_{\kappa}$ whenever $W$ is too large. We can estimate the maximum size of the sample from Eq. (\ref{eq:linear truncation}), where we find that $L_{\kappa}\approx 328$ nm. From this we can use Eq. (\ref{eq:Back Gate Truncation}) and equate it to the min$(L_{\kappa}, W)$ to find the minimum voltage needed to observe the effects of the back gate. 
	Samples such that $W\ll L_\kappa$ have been studied and their capacitance qualitatively agrees with our above predictions.\cite{Stemmer_concentration_interface}

	\section{Discussion}	
	\textit{Effective Mass}: In the above discussion we have assumed that the band structure of STO near the bottom of the conduction band consists of a single isotropic band with an effective mass $m^*$. In truth near the conduction
	band bottom of STO are three degenerate bands formed
	by $xy$, $xz$ and $yz$ Ti d-orbitals. This degeneracy is lifted by the spin-orbit interaction and results in two low energy bands that are nearly degenerate and a higher energy band offset by 20 meV\cite{Mazin_band_structure}. The mass $m^*$ used in the single band Thomas-Fermi approximation comes from the total density of states of all three bands at the Fermi surface. When $E_F\gg 20$ meV or equivalently $n>10^{19}$ cm$^{-3}$ all 3 bands contribute to $m^*$. At smaller concentrations, the high energy band is empty and no longer contributes to the density of states. This will slightly lower $m^*$. This minor difference in $m^*$ does not affect the dependence of the magnetization and specific heat on $W$, $\kappa$, and $V$ and instead only changes the parameter $b$ in all formulas.

	\textit{Rashba Interaction}: The Rashba spin-orbit interaction due to the breaking of inversion symmetry at the interface has been measured in LAO/STO gated structures. This interaction is characterized by the Rashba parameter $\alpha_R$ which is proportional to the electric field $E$\cite{cavigliarashba}. Near the surface where the field is largest it results in a splitting between bands by an amount $\Delta=2\alpha_R k_F\simeq 10$ meV at surface concentrations  $N=4.5\times10^{13}$ cm$^{-2}$. At such concentrations the Fermi energy $E_F(0)=18$ meV $\gtrsim10$ meV, so that near the surface the Rashba spin-orbit interaction is marginally small. Far from the surface in the tail of the electron density, which is most important for our results, $\Delta\ll E_F$. The reason for this is that even though the local Fermi energy $E_F(x)\propto x^{-8/7}$ at large $x$, the electric field $E\propto x^{-15/7}$ and $k_F\propto x^{-4/7}$. Therefore $\Delta/E_F\propto x^{-11/7}$ and the splitting quickly becomes irrelevant.
	
	\textit{Bulk Fermi level}: Above we have assumed that the bulk of STO is lightly doped by donors, so that the bulk Fermi level lies near the bottom of the conduction band and the density of electrons tends to zero at large $x$. Actually, bulk STO is believed to be heavily compensated so that Fermi level in the bulk is in the middle of the gap\cite{spinelli}. This does not affect the structure of the accumulation layer as the Fermi level does not acquire its bulk value until distances comparable to the screening radius of thermally activated carriers which is exponentially large at low temperatures.

	\section{Conclusion}

In this paper we have studied the thermodynamic properties of electron accumulation layers in STO created by positive charge at the surface. We have shown that the slow decay of the density profile causes divergence of the magnetization and specific heat per unit area at large distances from the surface. This anomalous behavior creates a dependence of these quantities on a truncation length and we have proposed several possible mechanisms for truncation. They lead to a nontrivial power law dependence of the magnetization and specific heat on the sample width $W$, the linear dielectric constant $\kappa$, or back gate voltage $V$. Additionally, we have studied the capacitance as a function of back gate voltage in thin samples where the tail of the electron gas has been partially depleted. The anomalous behavior of the magnetization, specific heat, and back gate capacitance due to the truncation of the tail allows for an experimental study of the shape of $n(x)$ and therefore a verification of the density profile given by Eq. (\ref{eq:nonlinear_concentration}). 

$\phantom{}$
\vspace*{2ex} \par \noindent
{\em Acknowledgments.}
We are grateful to K. V. Reich and V. Pribiag for helpful discussions. Han Fu was supported
by the Doctoral Dissertation Fellowship through the
University of Minnesota.

		\appendix
		\section{Calculation of the numerical constant $\gamma$ in the back gate truncation length}
		Eq. (\ref{eq:Back Gate Truncation}) was derived from a qualitative argument in which the electric fields from the back gate and the accumulation layer were matched. While this procedure should produce the correct scaling behavior, it is not reasonable to expect an accurate numerical coefficient in this way. In order to calculate the coefficient $\gamma$, we instead use the Thomas-Fermi approximation in which the self consistent potential $\varphi(x)$ is found to satisfy 
		\begin{equation}
		\frac{d}{dx}\left(\frac{d}{dx}\frac{\varphi}{e/b}\right)^{1/3}=\frac{2^{3/2}}{3\pi^2}\frac{A^{1/3}}{b^{4/3}}\left(\frac{a}{b}\right)^{4/3}\left(\frac{\varphi}{e/b}\right)^{3/2}\label{eq:Thomas Fermi},
		\end{equation}
		where $b=\hbar^2/m^*e^2\approx0.30$ $\AA$ has been introduced and $A=0.9$.\cite{RS} This equation was derived for the case of no back gate. In order to account for a back gate with voltage $-\abs{V}$applied to the sample, we simply change $\varphi\rightarrow\varphi-\abs{V}$ in Eq. (\ref{eq:Thomas Fermi})  In order to prepare for numerical calculations, it is useful to rewrite Eq. (\ref{eq:Thomas Fermi}) in a dimensionless form. Using $y=x/b$ and $\chi=(\varphi-\abs{V})/(e/b)$, we can write this as
		\begin{equation}
		\frac{d}{dy}\left(\frac{d\chi}{dy}\right)^{1/3}=\theta \chi^{3/2}\label{eq:chi},
		\end{equation}
		where
		 $\theta=2^{3/2}A^{1/3}(a/b)^{4/3}/(3\pi^2)$. It can be verified that this equation can be integrated to find
		\begin{equation}
		\frac{d\chi}{dy}=-\left(\frac{8}{5}\theta\chi^{5/2}+g_1\right)^{3/4}\label{eq:dimensionless Thomas Fermi}
		\end{equation}
		where $g_1$ is a constant of integration that can be related to the electric field at $L_V$ in the following way. We assume that the electrons only occupy a region $0<x<L_V$, and so the density profile $n(x)$ vanishes at $L_V$. Within the Thomas-Fermi approximation, we assume that the density is such that $\hbar^2(3\pi^2n)^{2/3}/2m^*=e(\varphi-\abs{V})$, from which it follows that $n(x)\propto\chi^{3/2}$. Therefore, $\chi$ must also vanish at $L_V$.  From this and Eq. (\ref{eq:dimensionless Thomas Fermi}), it immediately follows then that 
		\begin{equation}
		g_1=\left(\frac{d\chi}{dy}\right)^{4/3}\bigg\rvert_{L_V}.
		\end{equation}
		Because $\chi$ is the dimensionless form of the electric potential,  it follows that $-d\chi/dy$ is the electric field $E$ in units of $e/b^2$, and so $g_1=E^{4/3}$ in units of $e/b^2$. 
		
		Now that we understand the meaning of $g_1$, we can find $\gamma$ as follows. We first guess a value of $E(L_V)$. Once we make this guess, then we know both $g_1$ and the value of $\chi$ at $L_V$, and so $\chi$ is uniquely defined. Eq. (\ref{eq:dimensionless Thomas Fermi}) may then be used to numerically integrate from $L_V$ to any other value of $y$. In particular, we know what the value of the electric field is at $x=0$ where it must match the electric field of the positive charges near the surface given by (Eq. (\ref{eq:surface concentration})). So we may perform the integration until the value of $d\chi/dy=A(Nb^2)^3(a/b)^4$. If we track the change in $y$ during this procedure, we can find $L_V/b$ for this particular choice of $E(L_V)$. We can then repeat this process many times in order to generate a curve of $L_V$ vs. $E(L_V)$. Once this curve is obtained, we fit the data to the equation 
		\begin{equation}
		L_V=\gamma\left(\frac{b}{a}\right)^{8/15}E_V^{-7/15}
		\end{equation}
		where we have assumed the dependence of $L_V$ on $b/a$ and $E_V$ from Eq. (\ref{eq:Back Gate Truncation}) and $L_V$ and $E_V$ are in units of $b$ and $e/b^2$. Performing this procedure at a concentration $N=10^{14}$ cm$^{-2}$ gives us $\gamma=3.94$. Performing this at concentrations $N=3\times10^{14}$ cm$^{-2}$ does not change this value within the precision of our calculation. 
		
		\section{Table of coefficients}
		
		\begin{table}[h!]
			\begin{tabular}{|c|c|c|c|c|c|c|c|}
				\hline& & & & & & & \\[-8pt]
				$C_1$ & $C_2$ & $C_3$ & $C_4$ & $C_5$ & $C_6$ & $C_7$ & $A$\\
				\hline & & & & & & & \\[-8pt]
				5.8 & 1.3 & 2.4 & 2.1 & 3.9 & 3.7 & 3.7 & 0.5-1.5\\
				\hline
			\end{tabular}
			\caption{Values of the numerical coefficients for the Eqs. in the text.}\label{tab:coefficients}
		\end{table}
	\bibliography{papers22}

\begin{thebibliography}{39}%
\makeatletter
\providecommand \@ifxundefined [1]{%
 \@ifx{#1\undefined}
}%
\providecommand \@ifnum [1]{%
 \ifnum #1\expandafter \@firstoftwo
 \else \expandafter \@secondoftwo
 \fi
}%
\providecommand \@ifx [1]{%
 \ifx #1\expandafter \@firstoftwo
 \else \expandafter \@secondoftwo
 \fi
}%
\providecommand \natexlab [1]{#1}%
\providecommand \enquote  [1]{``#1''}%
\providecommand \bibnamefont  [1]{#1}%
\providecommand \bibfnamefont [1]{#1}%
\providecommand \citenamefont [1]{#1}%
\providecommand \href@noop [0]{\@secondoftwo}%
\providecommand \href [0]{\begingroup \@sanitize@url \@href}%
\providecommand \@href[1]{\@@startlink{#1}\@@href}%
\providecommand \@@href[1]{\endgroup#1\@@endlink}%
\providecommand \@sanitize@url [0]{\catcode `\\12\catcode `\$12\catcode
  `\&12\catcode `\#12\catcode `\^12\catcode `\_12\catcode `\%12\relax}%
\providecommand \@@startlink[1]{}%
\providecommand \@@endlink[0]{}%
\providecommand \url  [0]{\begingroup\@sanitize@url \@url }%
\providecommand \@url [1]{\endgroup\@href {#1}{\urlprefix }}%
\providecommand \urlprefix  [0]{URL }%
\providecommand \Eprint [0]{\href }%
\providecommand \doibase [0]{http://dx.doi.org/}%
\providecommand \selectlanguage [0]{\@gobble}%
\providecommand \bibinfo  [0]{\@secondoftwo}%
\providecommand \bibfield  [0]{\@secondoftwo}%
\providecommand \translation [1]{[#1]}%
\providecommand \BibitemOpen [0]{}%
\providecommand \bibitemStop [0]{}%
\providecommand \bibitemNoStop [0]{.\EOS\space}%
\providecommand \EOS [0]{\spacefactor3000\relax}%
\providecommand \BibitemShut  [1]{\csname bibitem#1\endcsname}%
\let\auto@bib@innerbib\@empty
\bibitem [{\citenamefont {Chakhalian}\ \emph {et~al.}(2014)\citenamefont
  {Chakhalian}, \citenamefont {Freeland}, \citenamefont {Millis}, \citenamefont
  {Panagopoulos},\ and\ \citenamefont {Rondinelli}}]{Oxides_rev}%
  \BibitemOpen
  \bibfield  {author} {\bibinfo {author} {\bibfnamefont {J.}~\bibnamefont
  {Chakhalian}}, \bibinfo {author} {\bibfnamefont {J.~W.}\ \bibnamefont
  {Freeland}}, \bibinfo {author} {\bibfnamefont {A.~J.}\ \bibnamefont
  {Millis}}, \bibinfo {author} {\bibfnamefont {C.}~\bibnamefont
  {Panagopoulos}}, \ and\ \bibinfo {author} {\bibfnamefont {J.~M.}\
  \bibnamefont {Rondinelli}},\ }\href {\doibase 10.1103/RevModPhys.86.1189}
  {\bibfield  {journal} {\bibinfo  {journal} {Rev. Mod. Phys.}\ }\textbf
  {\bibinfo {volume} {86}},\ \bibinfo {pages} {1189} (\bibinfo {year}
  {2014})}\BibitemShut {NoStop}%
\bibitem [{\citenamefont {Stemmer}\ and\ \citenamefont
  {Allen}(2014)}]{Stemmer_STO}%
  \BibitemOpen
  \bibfield  {author} {\bibinfo {author} {\bibfnamefont {S.}~\bibnamefont
  {Stemmer}}\ and\ \bibinfo {author} {\bibfnamefont {S.~J.}\ \bibnamefont
  {Allen}},\ }\href {\doibase 10.1146/annurev-matsci-070813-113552} {\bibfield
  {journal} {\bibinfo  {journal} {Annual Review of Materials Research}\
  }\textbf {\bibinfo {volume} {44}},\ \bibinfo {pages} {151} (\bibinfo {year}
  {2014})}\BibitemShut {NoStop}%
\bibitem [{\citenamefont {Zubko}\ \emph {et~al.}(2011)\citenamefont {Zubko},
  \citenamefont {Gariglio}, \citenamefont {Gabay}, \citenamefont {Ghosez},\
  and\ \citenamefont {Triscone}}]{Zubko_oxides}%
  \BibitemOpen
  \bibfield  {author} {\bibinfo {author} {\bibfnamefont {P.}~\bibnamefont
  {Zubko}}, \bibinfo {author} {\bibfnamefont {S.}~\bibnamefont {Gariglio}},
  \bibinfo {author} {\bibfnamefont {M.}~\bibnamefont {Gabay}}, \bibinfo
  {author} {\bibfnamefont {P.}~\bibnamefont {Ghosez}}, \ and\ \bibinfo {author}
  {\bibfnamefont {J.-M.}\ \bibnamefont {Triscone}},\ }\href {\doibase
  10.1146/annurev-conmatphys-062910-140445} {\bibfield  {journal} {\bibinfo
  {journal} {Annual Review of Condensed Matter Physics}\ }\textbf {\bibinfo
  {volume} {2}},\ \bibinfo {pages} {141} (\bibinfo {year} {2011})}\BibitemShut
  {NoStop}%
\bibitem [{\citenamefont {Noland}(1954)}]{Optical_absorbtion_STO}%
  \BibitemOpen
  \bibfield  {author} {\bibinfo {author} {\bibfnamefont {J.~A.}\ \bibnamefont
  {Noland}},\ }\href {\doibase 10.1103/PhysRev.94.724} {\bibfield  {journal}
  {\bibinfo  {journal} {Phys. Rev.}\ }\textbf {\bibinfo {volume} {94}},\
  \bibinfo {pages} {724} (\bibinfo {year} {1954})}\BibitemShut {NoStop}%
\bibitem [{\citenamefont {Ohtomo}\ and\ \citenamefont
  {Hwang}(2004)}]{Ohtomo_2004}%
  \BibitemOpen
  \bibfield  {author} {\bibinfo {author} {\bibfnamefont {A.}~\bibnamefont
  {Ohtomo}}\ and\ \bibinfo {author} {\bibfnamefont {H.~Y.}\ \bibnamefont
  {Hwang}},\ }\href {\doibase 10.1038/nature02308} {\bibfield  {journal}
  {\bibinfo  {journal} {Nature}\ }\textbf {\bibinfo {volume} {427}},\ \bibinfo
  {pages} {423} (\bibinfo {year} {2004})}\BibitemShut {NoStop}%
\bibitem [{\citenamefont {Xie}\ \emph {et~al.}(2013)\citenamefont {Xie},
  \citenamefont {Bell}, \citenamefont {Hikita}, \citenamefont {Harashima},\
  and\ \citenamefont {Hwang}}]{Hwang_mobility}%
  \BibitemOpen
  \bibfield  {author} {\bibinfo {author} {\bibfnamefont {Y.}~\bibnamefont
  {Xie}}, \bibinfo {author} {\bibfnamefont {C.}~\bibnamefont {Bell}}, \bibinfo
  {author} {\bibfnamefont {Y.}~\bibnamefont {Hikita}}, \bibinfo {author}
  {\bibfnamefont {S.}~\bibnamefont {Harashima}}, \ and\ \bibinfo {author}
  {\bibfnamefont {H.~Y.}\ \bibnamefont {Hwang}},\ }\href {\doibase
  10.1002/adma.201301798} {\bibfield  {journal} {\bibinfo  {journal} {Advanced
  Materials}\ }\textbf {\bibinfo {volume} {25}},\ \bibinfo {pages} {4735}
  (\bibinfo {year} {2013})}\BibitemShut {NoStop}%
\bibitem [{\citenamefont {Moetakef}\ \emph {et~al.}(2011)\citenamefont
  {Moetakef}, \citenamefont {Cain}, \citenamefont {Ouellette}, \citenamefont
  {Zhang}, \citenamefont {Klenov}, \citenamefont {Janotti}, \citenamefont
  {Van~de Walle}, \citenamefont {Rajan}, \citenamefont {Allen},\ and\
  \citenamefont {Stemmer}}]{Stemmer_GdTO}%
  \BibitemOpen
  \bibfield  {author} {\bibinfo {author} {\bibfnamefont {P.}~\bibnamefont
  {Moetakef}}, \bibinfo {author} {\bibfnamefont {T.~A.}\ \bibnamefont {Cain}},
  \bibinfo {author} {\bibfnamefont {D.~G.}\ \bibnamefont {Ouellette}}, \bibinfo
  {author} {\bibfnamefont {J.~Y.}\ \bibnamefont {Zhang}}, \bibinfo {author}
  {\bibfnamefont {D.~O.}\ \bibnamefont {Klenov}}, \bibinfo {author}
  {\bibfnamefont {A.}~\bibnamefont {Janotti}}, \bibinfo {author} {\bibfnamefont
  {C.~G.}\ \bibnamefont {Van~de Walle}}, \bibinfo {author} {\bibfnamefont
  {S.}~\bibnamefont {Rajan}}, \bibinfo {author} {\bibfnamefont {S.~J.}\
  \bibnamefont {Allen}}, \ and\ \bibinfo {author} {\bibfnamefont
  {S.}~\bibnamefont {Stemmer}},\ }\href {\doibase 10.1063/1.3669402} {\bibfield
   {journal} {\bibinfo  {journal} {Applied Physics Letters}\ }\textbf {\bibinfo
  {volume} {99}},\ \bibinfo {pages} {232116} (\bibinfo {year}
  {2011})}\BibitemShut {NoStop}%
\bibitem [{\citenamefont {He}\ \emph {et~al.}(2012)\citenamefont {He},
  \citenamefont {Sanders}, \citenamefont {Gray}, \citenamefont {Wong},
  \citenamefont {Mehta},\ and\ \citenamefont {Suzuki}}]{LaVO_STO}%
  \BibitemOpen
  \bibfield  {author} {\bibinfo {author} {\bibfnamefont {C.}~\bibnamefont
  {He}}, \bibinfo {author} {\bibfnamefont {T.~D.}\ \bibnamefont {Sanders}},
  \bibinfo {author} {\bibfnamefont {M.~T.}\ \bibnamefont {Gray}}, \bibinfo
  {author} {\bibfnamefont {F.~J.}\ \bibnamefont {Wong}}, \bibinfo {author}
  {\bibfnamefont {V.~V.}\ \bibnamefont {Mehta}}, \ and\ \bibinfo {author}
  {\bibfnamefont {Y.}~\bibnamefont {Suzuki}},\ }\href {\doibase
  10.1103/PhysRevB.86.081401} {\bibfield  {journal} {\bibinfo  {journal} {Phys.
  Rev. B}\ }\textbf {\bibinfo {volume} {86}},\ \bibinfo {pages} {081401}
  (\bibinfo {year} {2012})}\BibitemShut {NoStop}%
\bibitem [{\citenamefont {Annadi}\ \emph {et~al.}(2012)\citenamefont {Annadi},
  \citenamefont {Putra}, \citenamefont {Liu}, \citenamefont {Wang},
  \citenamefont {Gopinadhan}, \citenamefont {Huang}, \citenamefont {Dhar},
  \citenamefont {Venkatesan},\ and\ \citenamefont
  {Ariando}}]{different_polar_STO}%
  \BibitemOpen
  \bibfield  {author} {\bibinfo {author} {\bibfnamefont {A.}~\bibnamefont
  {Annadi}}, \bibinfo {author} {\bibfnamefont {A.}~\bibnamefont {Putra}},
  \bibinfo {author} {\bibfnamefont {Z.~Q.}\ \bibnamefont {Liu}}, \bibinfo
  {author} {\bibfnamefont {X.}~\bibnamefont {Wang}}, \bibinfo {author}
  {\bibfnamefont {K.}~\bibnamefont {Gopinadhan}}, \bibinfo {author}
  {\bibfnamefont {Z.}~\bibnamefont {Huang}}, \bibinfo {author} {\bibfnamefont
  {S.}~\bibnamefont {Dhar}}, \bibinfo {author} {\bibfnamefont {T.}~\bibnamefont
  {Venkatesan}}, \ and\ \bibinfo {author} {\bibnamefont {Ariando}},\ }\href
  {\doibase 10.1103/PhysRevB.86.085450} {\bibfield  {journal} {\bibinfo
  {journal} {Phys. Rev. B}\ }\textbf {\bibinfo {volume} {86}},\ \bibinfo
  {pages} {085450} (\bibinfo {year} {2012})}\BibitemShut {NoStop}%
\bibitem [{\citenamefont {Perna}\ \emph {et~al.}(2010)\citenamefont {Perna},
  \citenamefont {Maccariello}, \citenamefont {Radovic}, \citenamefont
  {Scotti~di Uccio}, \citenamefont {Pallecchi}, \citenamefont {Codda},
  \citenamefont {Marré}, \citenamefont {Cantoni}, \citenamefont {Gazquez},
  \citenamefont {Varela}, \citenamefont {Pennycook},\ and\ \citenamefont
  {Granozio}}]{LaGaO_STO}%
  \BibitemOpen
  \bibfield  {author} {\bibinfo {author} {\bibfnamefont {P.}~\bibnamefont
  {Perna}}, \bibinfo {author} {\bibfnamefont {D.}~\bibnamefont {Maccariello}},
  \bibinfo {author} {\bibfnamefont {M.}~\bibnamefont {Radovic}}, \bibinfo
  {author} {\bibfnamefont {U.}~\bibnamefont {Scotti~di Uccio}}, \bibinfo
  {author} {\bibfnamefont {I.}~\bibnamefont {Pallecchi}}, \bibinfo {author}
  {\bibfnamefont {M.}~\bibnamefont {Codda}}, \bibinfo {author} {\bibfnamefont
  {D.}~\bibnamefont {Marré}}, \bibinfo {author} {\bibfnamefont
  {C.}~\bibnamefont {Cantoni}}, \bibinfo {author} {\bibfnamefont
  {J.}~\bibnamefont {Gazquez}}, \bibinfo {author} {\bibfnamefont
  {M.}~\bibnamefont {Varela}}, \bibinfo {author} {\bibfnamefont {S.~J.}\
  \bibnamefont {Pennycook}}, \ and\ \bibinfo {author} {\bibfnamefont {F.~M.}\
  \bibnamefont {Granozio}},\ }\href {\doibase 10.1063/1.3496440} {\bibfield
  {journal} {\bibinfo  {journal} {Applied Physics Letters}\ }\textbf {\bibinfo
  {volume} {97}},\ \bibinfo {pages} {152111} (\bibinfo {year}
  {2010})}\BibitemShut {NoStop}%
\bibitem [{\citenamefont {Ohtsuka}\ \emph {et~al.}(2010)\citenamefont
  {Ohtsuka}, \citenamefont {Matvejeff}, \citenamefont {Nishio}, \citenamefont
  {Takahashi},\ and\ \citenamefont {Lippmaa}}]{LaTO_STO}%
  \BibitemOpen
  \bibfield  {author} {\bibinfo {author} {\bibfnamefont {R.}~\bibnamefont
  {Ohtsuka}}, \bibinfo {author} {\bibfnamefont {M.}~\bibnamefont {Matvejeff}},
  \bibinfo {author} {\bibfnamefont {K.}~\bibnamefont {Nishio}}, \bibinfo
  {author} {\bibfnamefont {R.}~\bibnamefont {Takahashi}}, \ and\ \bibinfo
  {author} {\bibfnamefont {M.}~\bibnamefont {Lippmaa}},\ }\href {\doibase
  10.1063/1.3430006} {\bibfield  {journal} {\bibinfo  {journal} {Applied
  Physics Letters}\ }\textbf {\bibinfo {volume} {96}},\ \bibinfo {pages}
  {192111} (\bibinfo {year} {2010})}\BibitemShut {NoStop}%
\bibitem [{\citenamefont {Thiel}\ \emph {et~al.}(2006)\citenamefont {Thiel},
  \citenamefont {Hammerl}, \citenamefont {Schmehl}, \citenamefont {Schneider},\
  and\ \citenamefont {Mannhart}}]{10_percent}%
  \BibitemOpen
  \bibfield  {author} {\bibinfo {author} {\bibfnamefont {S.}~\bibnamefont
  {Thiel}}, \bibinfo {author} {\bibfnamefont {G.}~\bibnamefont {Hammerl}},
  \bibinfo {author} {\bibfnamefont {A.}~\bibnamefont {Schmehl}}, \bibinfo
  {author} {\bibfnamefont {C.~W.}\ \bibnamefont {Schneider}}, \ and\ \bibinfo
  {author} {\bibfnamefont {J.}~\bibnamefont {Mannhart}},\ }\href {\doibase
  10.1126/science.1131091} {\bibfield  {journal} {\bibinfo  {journal}
  {Science}\ }\textbf {\bibinfo {volume} {313}},\ \bibinfo {pages} {1942}
  (\bibinfo {year} {2006})}\BibitemShut {NoStop}%
\bibitem [{\citenamefont {Hosoda}\ \emph {et~al.}(2013)\citenamefont {Hosoda},
  \citenamefont {Bell}, \citenamefont {Hikita},\ and\ \citenamefont
  {Hwang}}]{Hwang_gate}%
  \BibitemOpen
  \bibfield  {author} {\bibinfo {author} {\bibfnamefont {M.}~\bibnamefont
  {Hosoda}}, \bibinfo {author} {\bibfnamefont {C.}~\bibnamefont {Bell}},
  \bibinfo {author} {\bibfnamefont {Y.}~\bibnamefont {Hikita}}, \ and\ \bibinfo
  {author} {\bibfnamefont {H.~Y.}\ \bibnamefont {Hwang}},\ }\href {\doibase
  10.1063/1.4794410} {\bibfield  {journal} {\bibinfo  {journal} {Applied
  Physics Letters}\ }\textbf {\bibinfo {volume} {102}},\ \bibinfo {eid}
  {091601} (\bibinfo {year} {2013})}\BibitemShut {NoStop}%
\bibitem [{\citenamefont {Boucherit}\ \emph {et~al.}(2014)\citenamefont
  {Boucherit}, \citenamefont {Shoron}, \citenamefont {Jackson}, \citenamefont
  {Cain}, \citenamefont {Buffon}, \citenamefont {Polchinski}, \citenamefont
  {Stemmer},\ and\ \citenamefont {Rajan}}]{Stemmer_concentration_interface}%
  \BibitemOpen
  \bibfield  {author} {\bibinfo {author} {\bibfnamefont {M.}~\bibnamefont
  {Boucherit}}, \bibinfo {author} {\bibfnamefont {O.}~\bibnamefont {Shoron}},
  \bibinfo {author} {\bibfnamefont {C.~A.}\ \bibnamefont {Jackson}}, \bibinfo
  {author} {\bibfnamefont {T.~A.}\ \bibnamefont {Cain}}, \bibinfo {author}
  {\bibfnamefont {M.~L.~C.}\ \bibnamefont {Buffon}}, \bibinfo {author}
  {\bibfnamefont {C.}~\bibnamefont {Polchinski}}, \bibinfo {author}
  {\bibfnamefont {S.}~\bibnamefont {Stemmer}}, \ and\ \bibinfo {author}
  {\bibfnamefont {S.}~\bibnamefont {Rajan}},\ }\href {\doibase
  10.1063/1.4875796} {\bibfield  {journal} {\bibinfo  {journal} {Applied
  Physics Letters}\ }\textbf {\bibinfo {volume} {104}},\ \bibinfo {eid}
  {182904} (\bibinfo {year} {2014})}\BibitemShut {NoStop}%
\bibitem [{\citenamefont {Ueno}\ \emph {et~al.}(2008)\citenamefont {Ueno},
  \citenamefont {Nakamura}, \citenamefont {Shimotani}, \citenamefont {Ohtomo},
  \citenamefont {Kimura}, \citenamefont {Nojima}, \citenamefont {Aoki},
  \citenamefont {Iwasa},\ and\ \citenamefont
  {Kawasaki}}]{induced_superconductivity}%
  \BibitemOpen
  \bibfield  {author} {\bibinfo {author} {\bibfnamefont {K.}~\bibnamefont
  {Ueno}}, \bibinfo {author} {\bibfnamefont {S.}~\bibnamefont {Nakamura}},
  \bibinfo {author} {\bibfnamefont {H.}~\bibnamefont {Shimotani}}, \bibinfo
  {author} {\bibfnamefont {A.}~\bibnamefont {Ohtomo}}, \bibinfo {author}
  {\bibfnamefont {N.}~\bibnamefont {Kimura}}, \bibinfo {author} {\bibfnamefont
  {T.}~\bibnamefont {Nojima}}, \bibinfo {author} {\bibfnamefont
  {H.}~\bibnamefont {Aoki}}, \bibinfo {author} {\bibfnamefont {Y.}~\bibnamefont
  {Iwasa}}, \ and\ \bibinfo {author} {\bibfnamefont {M.}~\bibnamefont
  {Kawasaki}},\ }\href {\doibase 10.1038/nmat2298} {\bibfield  {journal}
  {\bibinfo  {journal} {Nat Mater}\ }\textbf {\bibinfo {volume} {7}},\ \bibinfo
  {pages} {855} (\bibinfo {year} {2008})}\BibitemShut {NoStop}%
\bibitem [{\citenamefont {Gallagher}\ \emph {et~al.}(2014)\citenamefont
  {Gallagher}, \citenamefont {Lee}, \citenamefont {Williams},\ and\
  \citenamefont {Goldhaber-Gordon}}]{Gallagher_2014}%
  \BibitemOpen
  \bibfield  {author} {\bibinfo {author} {\bibfnamefont {P.}~\bibnamefont
  {Gallagher}}, \bibinfo {author} {\bibfnamefont {M.}~\bibnamefont {Lee}},
  \bibinfo {author} {\bibfnamefont {J.~R.}\ \bibnamefont {Williams}}, \ and\
  \bibinfo {author} {\bibfnamefont {D.}~\bibnamefont {Goldhaber-Gordon}},\
  }\href {\doibase 10.1038/nphys3049} {\bibfield  {journal} {\bibinfo
  {journal} {Nature Physics}\ }\textbf {\bibinfo {volume} {10}},\ \bibinfo
  {pages} {748} (\bibinfo {year} {2014})}\BibitemShut {NoStop}%
\bibitem [{\citenamefont {Jalan}\ \emph {et~al.}(2010)\citenamefont {Jalan},
  \citenamefont {Stemmer}, \citenamefont {Mack},\ and\ \citenamefont
  {Allen}}]{delta_doped_stemmer}%
  \BibitemOpen
  \bibfield  {author} {\bibinfo {author} {\bibfnamefont {B.}~\bibnamefont
  {Jalan}}, \bibinfo {author} {\bibfnamefont {S.}~\bibnamefont {Stemmer}},
  \bibinfo {author} {\bibfnamefont {S.}~\bibnamefont {Mack}}, \ and\ \bibinfo
  {author} {\bibfnamefont {S.~J.}\ \bibnamefont {Allen}},\ }\href {\doibase
  10.1103/PhysRevB.82.081103} {\bibfield  {journal} {\bibinfo  {journal} {Phys.
  Rev. B}\ }\textbf {\bibinfo {volume} {82}},\ \bibinfo {pages} {081103}
  (\bibinfo {year} {2010})}\BibitemShut {NoStop}%
\bibitem [{\citenamefont {Kozuka}\ \emph {et~al.}(2010)\citenamefont {Kozuka},
  \citenamefont {Kim}, \citenamefont {Ohta}, \citenamefont {Hikita},
  \citenamefont {Bell},\ and\ \citenamefont {Hwang}}]{delta_doped_STO_Hwang}%
  \BibitemOpen
  \bibfield  {author} {\bibinfo {author} {\bibfnamefont {Y.}~\bibnamefont
  {Kozuka}}, \bibinfo {author} {\bibfnamefont {M.}~\bibnamefont {Kim}},
  \bibinfo {author} {\bibfnamefont {H.}~\bibnamefont {Ohta}}, \bibinfo {author}
  {\bibfnamefont {Y.}~\bibnamefont {Hikita}}, \bibinfo {author} {\bibfnamefont
  {C.}~\bibnamefont {Bell}}, \ and\ \bibinfo {author} {\bibfnamefont {H.~Y.}\
  \bibnamefont {Hwang}},\ }\href {\doibase 10.1063/1.3524198} {\bibfield
  {journal} {\bibinfo  {journal} {Applied Physics Letters}\ }\textbf {\bibinfo
  {volume} {97}},\ \bibinfo {eid} {222115} (\bibinfo {year}
  {2010})}\BibitemShut {NoStop}%
\bibitem [{\citenamefont {Kaiser}\ \emph {et~al.}(2012)\citenamefont {Kaiser},
  \citenamefont {Gray}, \citenamefont {Conti}, \citenamefont {Jalan},
  \citenamefont {Kajdos}, \citenamefont {Gloskovskii}, \citenamefont {Ueda},
  \citenamefont {Yamashita}, \citenamefont {Kobayashi}, \citenamefont {Drube},
  \citenamefont {Stemmer},\ and\ \citenamefont
  {Fadley}}]{delta_doped_STO_Stemmer}%
  \BibitemOpen
  \bibfield  {author} {\bibinfo {author} {\bibfnamefont {A.~M.}\ \bibnamefont
  {Kaiser}}, \bibinfo {author} {\bibfnamefont {A.~X.}\ \bibnamefont {Gray}},
  \bibinfo {author} {\bibfnamefont {G.}~\bibnamefont {Conti}}, \bibinfo
  {author} {\bibfnamefont {B.}~\bibnamefont {Jalan}}, \bibinfo {author}
  {\bibfnamefont {A.~P.}\ \bibnamefont {Kajdos}}, \bibinfo {author}
  {\bibfnamefont {A.}~\bibnamefont {Gloskovskii}}, \bibinfo {author}
  {\bibfnamefont {S.}~\bibnamefont {Ueda}}, \bibinfo {author} {\bibfnamefont
  {Y.}~\bibnamefont {Yamashita}}, \bibinfo {author} {\bibfnamefont
  {K.}~\bibnamefont {Kobayashi}}, \bibinfo {author} {\bibfnamefont
  {W.}~\bibnamefont {Drube}}, \bibinfo {author} {\bibfnamefont
  {S.}~\bibnamefont {Stemmer}}, \ and\ \bibinfo {author} {\bibfnamefont
  {C.~S.}\ \bibnamefont {Fadley}},\ }\href {\doibase 10.1063/1.4731642}
  {\bibfield  {journal} {\bibinfo  {journal} {Applied Physics Letters}\
  }\textbf {\bibinfo {volume} {100}},\ \bibinfo {eid} {261603} (\bibinfo {year}
  {2012})}\BibitemShut {NoStop}%
\bibitem [{\citenamefont {Minohara}\ \emph {et~al.}(2014)\citenamefont
  {Minohara}, \citenamefont {Hikita}, \citenamefont {Bell}, \citenamefont
  {Inoue}, \citenamefont {Hosoda}, \citenamefont {Sato}, \citenamefont
  {Kumigashira}, \citenamefont {Oshima}, \citenamefont {Ikenaga},\ and\
  \citenamefont {Hwang}}]{Hwang_Xray}%
  \BibitemOpen
  \bibfield  {author} {\bibinfo {author} {\bibfnamefont {M.}~\bibnamefont
  {Minohara}}, \bibinfo {author} {\bibfnamefont {Y.}~\bibnamefont {Hikita}},
  \bibinfo {author} {\bibfnamefont {C.}~\bibnamefont {Bell}}, \bibinfo {author}
  {\bibfnamefont {H.}~\bibnamefont {Inoue}}, \bibinfo {author} {\bibfnamefont
  {M.}~\bibnamefont {Hosoda}}, \bibinfo {author} {\bibfnamefont {H.~K.}\
  \bibnamefont {Sato}}, \bibinfo {author} {\bibfnamefont {H.}~\bibnamefont
  {Kumigashira}}, \bibinfo {author} {\bibfnamefont {M.}~\bibnamefont {Oshima}},
  \bibinfo {author} {\bibfnamefont {E.}~\bibnamefont {Ikenaga}}, \ and\
  \bibinfo {author} {\bibfnamefont {H.~Y.}\ \bibnamefont {Hwang}},\ }\href
  {http://arxiv.org/abs/1403.5594} {\bibfield  {journal} {\bibinfo  {journal}
  {arXiv:1403.5594}\ } (\bibinfo {year} {2014})}\BibitemShut {NoStop}%
\bibitem [{\citenamefont {Yamada}\ \emph {et~al.}(2014)\citenamefont {Yamada},
  \citenamefont {Sato}, \citenamefont {Hikita}, \citenamefont {Hwang},\ and\
  \citenamefont {Kanemitsu}}]{Hwang_PL}%
  \BibitemOpen
  \bibfield  {author} {\bibinfo {author} {\bibfnamefont {Y.}~\bibnamefont
  {Yamada}}, \bibinfo {author} {\bibfnamefont {H.~K.}\ \bibnamefont {Sato}},
  \bibinfo {author} {\bibfnamefont {Y.}~\bibnamefont {Hikita}}, \bibinfo
  {author} {\bibfnamefont {H.~Y.}\ \bibnamefont {Hwang}}, \ and\ \bibinfo
  {author} {\bibfnamefont {Y.}~\bibnamefont {Kanemitsu}},\ }\href {\doibase
  10.1063/1.4872171} {\bibfield  {journal} {\bibinfo  {journal} {Applied
  Physics Letters}\ }\textbf {\bibinfo {volume} {104}},\ \bibinfo {pages}
  {151907} (\bibinfo {year} {2014})}\BibitemShut {NoStop}%
\bibitem [{\citenamefont {Dubroka}\ \emph {et~al.}(2010)\citenamefont
  {Dubroka}, \citenamefont {R\"ossle}, \citenamefont {Kim}, \citenamefont
  {Malik}, \citenamefont {Schultz}, \citenamefont {Thiel}, \citenamefont
  {Schneider}, \citenamefont {Mannhart}, \citenamefont {Herranz}, \citenamefont
  {Copie}, \citenamefont {Bibes}, \citenamefont {Barth\'el\'emy},\ and\
  \citenamefont {Bernhard}}]{LAO_STO_Berreman}%
  \BibitemOpen
  \bibfield  {author} {\bibinfo {author} {\bibfnamefont {A.}~\bibnamefont
  {Dubroka}}, \bibinfo {author} {\bibfnamefont {M.}~\bibnamefont {R\"ossle}},
  \bibinfo {author} {\bibfnamefont {K.~W.}\ \bibnamefont {Kim}}, \bibinfo
  {author} {\bibfnamefont {V.~K.}\ \bibnamefont {Malik}}, \bibinfo {author}
  {\bibfnamefont {L.}~\bibnamefont {Schultz}}, \bibinfo {author} {\bibfnamefont
  {S.}~\bibnamefont {Thiel}}, \bibinfo {author} {\bibfnamefont {C.~W.}\
  \bibnamefont {Schneider}}, \bibinfo {author} {\bibfnamefont {J.}~\bibnamefont
  {Mannhart}}, \bibinfo {author} {\bibfnamefont {G.}~\bibnamefont {Herranz}},
  \bibinfo {author} {\bibfnamefont {O.}~\bibnamefont {Copie}}, \bibinfo
  {author} {\bibfnamefont {M.}~\bibnamefont {Bibes}}, \bibinfo {author}
  {\bibfnamefont {A.}~\bibnamefont {Barth\'el\'emy}}, \ and\ \bibinfo {author}
  {\bibfnamefont {C.}~\bibnamefont {Bernhard}},\ }\href {\doibase
  10.1103/PhysRevLett.104.156807} {\bibfield  {journal} {\bibinfo  {journal}
  {Phys. Rev. Lett.}\ }\textbf {\bibinfo {volume} {104}},\ \bibinfo {pages}
  {156807} (\bibinfo {year} {2010})}\BibitemShut {NoStop}%
\bibitem [{\citenamefont {Khalsa}\ and\ \citenamefont
  {MacDonald}(2012)}]{MacDonald_theory}%
  \BibitemOpen
  \bibfield  {author} {\bibinfo {author} {\bibfnamefont {G.}~\bibnamefont
  {Khalsa}}\ and\ \bibinfo {author} {\bibfnamefont {A.~H.}\ \bibnamefont
  {MacDonald}},\ }\href {\doibase 10.1103/PhysRevB.86.125121} {\bibfield
  {journal} {\bibinfo  {journal} {Phys. Rev. B}\ }\textbf {\bibinfo {volume}
  {86}},\ \bibinfo {pages} {125121} (\bibinfo {year} {2012})}\BibitemShut
  {NoStop}%
\bibitem [{\citenamefont {Stengel}(2011)}]{abinitio_STO}%
  \BibitemOpen
  \bibfield  {author} {\bibinfo {author} {\bibfnamefont {M.}~\bibnamefont
  {Stengel}},\ }\href {\doibase 10.1103/PhysRevLett.106.136803} {\bibfield
  {journal} {\bibinfo  {journal} {Phys. Rev. Lett.}\ }\textbf {\bibinfo
  {volume} {106}},\ \bibinfo {pages} {136803} (\bibinfo {year}
  {2011})}\BibitemShut {NoStop}%
\bibitem [{\citenamefont {Son}\ \emph {et~al.}(2009)\citenamefont {Son},
  \citenamefont {Cho}, \citenamefont {Lee}, \citenamefont {Lee},\ and\
  \citenamefont {Han}}]{abinitio_STO_2}%
  \BibitemOpen
  \bibfield  {author} {\bibinfo {author} {\bibfnamefont {W.-j.}\ \bibnamefont
  {Son}}, \bibinfo {author} {\bibfnamefont {E.}~\bibnamefont {Cho}}, \bibinfo
  {author} {\bibfnamefont {B.}~\bibnamefont {Lee}}, \bibinfo {author}
  {\bibfnamefont {J.}~\bibnamefont {Lee}}, \ and\ \bibinfo {author}
  {\bibfnamefont {S.}~\bibnamefont {Han}},\ }\href {\doibase
  10.1103/PhysRevB.79.245411} {\bibfield  {journal} {\bibinfo  {journal} {Phys.
  Rev. B}\ }\textbf {\bibinfo {volume} {79}},\ \bibinfo {pages} {245411}
  (\bibinfo {year} {2009})}\BibitemShut {NoStop}%
\bibitem [{\citenamefont {Park}\ and\ \citenamefont
  {Millis}(2013)}]{distribution_LAO_STO}%
  \BibitemOpen
  \bibfield  {author} {\bibinfo {author} {\bibfnamefont {S.~Y.}\ \bibnamefont
  {Park}}\ and\ \bibinfo {author} {\bibfnamefont {A.~J.}\ \bibnamefont
  {Millis}},\ }\href {\doibase 10.1103/PhysRevB.87.205145} {\bibfield
  {journal} {\bibinfo  {journal} {Phys. Rev. B}\ }\textbf {\bibinfo {volume}
  {87}},\ \bibinfo {pages} {205145} (\bibinfo {year} {2013})}\BibitemShut
  {NoStop}%
\bibitem [{\citenamefont {Haraldsen}\ \emph {et~al.}(2012)\citenamefont
  {Haraldsen}, \citenamefont {W\"olfle},\ and\ \citenamefont
  {Balatsky}}]{superconductivity_LAO_STO}%
  \BibitemOpen
  \bibfield  {author} {\bibinfo {author} {\bibfnamefont {J.~T.}\ \bibnamefont
  {Haraldsen}}, \bibinfo {author} {\bibfnamefont {P.}~\bibnamefont {W\"olfle}},
  \ and\ \bibinfo {author} {\bibfnamefont {A.~V.}\ \bibnamefont {Balatsky}},\
  }\href {\doibase 10.1103/PhysRevB.85.134501} {\bibfield  {journal} {\bibinfo
  {journal} {Phys. Rev. B}\ }\textbf {\bibinfo {volume} {85}},\ \bibinfo
  {pages} {134501} (\bibinfo {year} {2012})}\BibitemShut {NoStop}%
\bibitem [{\citenamefont {Reich}\ \emph {et~al.}(2015)\citenamefont {Reich},
  \citenamefont {Schecter},\ and\ \citenamefont {Shklovskii}}]{RS}%
  \BibitemOpen
  \bibfield  {author} {\bibinfo {author} {\bibfnamefont {K.~V.}\ \bibnamefont
  {Reich}}, \bibinfo {author} {\bibfnamefont {M.}~\bibnamefont {Schecter}}, \
  and\ \bibinfo {author} {\bibfnamefont {B.~I.}\ \bibnamefont {Shklovskii}},\
  }\href {\doibase 10.1103/PhysRevB.91.115303} {\bibfield  {journal} {\bibinfo
  {journal} {Phys. Rev. B}\ }\textbf {\bibinfo {volume} {91}},\ \bibinfo
  {pages} {115303} (\bibinfo {year} {2015})}\BibitemShut {NoStop}%
\bibitem [{\citenamefont {Ginzburg}(1946)}]{Ginzburg_ferroelectrics}%
  \BibitemOpen
  \bibfield  {author} {\bibinfo {author} {\bibfnamefont {V.}~\bibnamefont
  {Ginzburg}},\ }\href@noop {} {\bibfield  {journal} {\bibinfo  {journal} {J.
  Phys. USSR}\ }\textbf {\bibinfo {volume} {10}},\ \bibinfo {pages} {107}
  (\bibinfo {year} {1946})}\BibitemShut {NoStop}%
\bibitem [{\citenamefont {Landau}\ and\ \citenamefont
  {Lifshitz}(1980)}]{Landau_stat}%
  \BibitemOpen
  \bibfield  {author} {\bibinfo {author} {\bibfnamefont {L.~D.}\ \bibnamefont
  {Landau}}\ and\ \bibinfo {author} {\bibfnamefont {E.~M.}\ \bibnamefont
  {Lifshitz}},\ }\href@noop {} {\emph {\bibinfo {title} {Statistical Mechanics
  (Part 1)}}},\ edited by\ \bibinfo {editor} {\bibfnamefont {E.~M.}\
  \bibnamefont {Lifshitz}}\ and\ \bibinfo {editor} {\bibfnamefont {L.~P.}\
  \bibnamefont {Pitaevskii}},\ \bibinfo {series} {Course of Theoretical
  Physics}, Vol.~\bibinfo {volume} {5}\ (\bibinfo  {publisher}
  {Butterworth-Heinemann},\ \bibinfo {year} {1980})\BibitemShut {NoStop}%
\bibitem [{\citenamefont {Thomas}(1927)}]{Thomas_Fermi}%
  \BibitemOpen
  \bibfield  {author} {\bibinfo {author} {\bibfnamefont {L.~H.}\ \bibnamefont
  {Thomas}},\ }\href {\doibase 10.1017/S0305004100011683} {\bibfield  {journal}
  {\bibinfo  {journal} {Mathematical Proceedings of the Cambridge Philosophical
  Society}\ }\textbf {\bibinfo {volume} {23}},\ \bibinfo {pages} {542}
  (\bibinfo {year} {1927})}\BibitemShut {NoStop}%
\bibitem [{\citenamefont {Fu}\ \emph {et~al.}(2016{\natexlab{a}})\citenamefont
  {Fu}, \citenamefont {Reich},\ and\ \citenamefont {Shklovskii}}]{RAT}%
  \BibitemOpen
  \bibfield  {author} {\bibinfo {author} {\bibfnamefont {H.}~\bibnamefont
  {Fu}}, \bibinfo {author} {\bibfnamefont {K.}~\bibnamefont {Reich}}, \ and\
  \bibinfo {author} {\bibfnamefont {B.}~\bibnamefont {Shklovskii}},\
  }\href@noop {} {\bibfield  {journal} {\bibinfo  {journal} {Physical Review
  B}\ }\textbf {\bibinfo {volume} {94}},\ \bibinfo {pages} {045310} (\bibinfo
  {year} {2016}{\natexlab{a}})}\BibitemShut {NoStop}%
\bibitem [{\citenamefont {Fu}\ \emph {et~al.}(2016{\natexlab{b}})\citenamefont
  {Fu}, \citenamefont {Reich},\ and\ \citenamefont {Shklovskii}}]{HanSTO}%
  \BibitemOpen
  \bibfield  {author} {\bibinfo {author} {\bibfnamefont {H.}~\bibnamefont
  {Fu}}, \bibinfo {author} {\bibfnamefont {K.}~\bibnamefont {Reich}}, \ and\
  \bibinfo {author} {\bibfnamefont {B.}~\bibnamefont {Shklovskii}},\
  }\href@noop {} {\bibfield  {journal} {\bibinfo  {journal} {Journal of
  Experimental and Theoretical Physics}\ }\textbf {\bibinfo {volume} {122}},\
  \bibinfo {pages} {456} (\bibinfo {year} {2016}{\natexlab{b}})}\BibitemShut
  {NoStop}%
\bibitem [{Note1()}]{Note1}%
  \BibitemOpen
  \bibinfo {note} {When $V>0$ extra electrons are brought to STO. They form an
  accumulation layer at the interface with a gate, similar to the symmetric
  wells GTO/STO/GTO As a result the magnetization and specific heat will grow
  with $V$ and in the symmetric case are enhanced by a factor $2^{4/7}$
  compared to when $V=0$ and the truncation was by the finite sample width
  $W$.}\BibitemShut {Stop}%
\bibitem [{\citenamefont {Li}\ \emph {et~al.}(2011)\citenamefont {Li},
  \citenamefont {Richter}, \citenamefont {Paetel}, \citenamefont {Kopp},
  \citenamefont {Mannhart},\ and\ \citenamefont {Ashoori}}]{licapacitance}%
  \BibitemOpen
  \bibfield  {author} {\bibinfo {author} {\bibfnamefont {L.}~\bibnamefont
  {Li}}, \bibinfo {author} {\bibfnamefont {C.}~\bibnamefont {Richter}},
  \bibinfo {author} {\bibfnamefont {S.}~\bibnamefont {Paetel}}, \bibinfo
  {author} {\bibfnamefont {T.}~\bibnamefont {Kopp}}, \bibinfo {author}
  {\bibfnamefont {J.}~\bibnamefont {Mannhart}}, \ and\ \bibinfo {author}
  {\bibfnamefont {R.}~\bibnamefont {Ashoori}},\ }\href@noop {} {\bibfield
  {journal} {\bibinfo  {journal} {Science}\ }\textbf {\bibinfo {volume}
  {332}},\ \bibinfo {pages} {825} (\bibinfo {year} {2011})}\BibitemShut
  {NoStop}%
\bibitem [{\citenamefont {Tinkl}\ \emph {et~al.}(2012)\citenamefont {Tinkl},
  \citenamefont {Breitschaft}, \citenamefont {Richter},\ and\ \citenamefont
  {Mannhart}}]{tinkl}%
  \BibitemOpen
  \bibfield  {author} {\bibinfo {author} {\bibfnamefont {V.}~\bibnamefont
  {Tinkl}}, \bibinfo {author} {\bibfnamefont {M.}~\bibnamefont {Breitschaft}},
  \bibinfo {author} {\bibfnamefont {C.}~\bibnamefont {Richter}}, \ and\
  \bibinfo {author} {\bibfnamefont {J.}~\bibnamefont {Mannhart}},\ }\href@noop
  {} {\bibfield  {journal} {\bibinfo  {journal} {Physical Review B}\ }\textbf
  {\bibinfo {volume} {86}},\ \bibinfo {pages} {075116} (\bibinfo {year}
  {2012})}\BibitemShut {NoStop}%
\bibitem [{\citenamefont {van~der Marel}\ \emph {et~al.}(2011)\citenamefont
  {van~der Marel}, \citenamefont {van Mechelen},\ and\ \citenamefont
  {Mazin}}]{Mazin_band_structure}%
  \BibitemOpen
  \bibfield  {author} {\bibinfo {author} {\bibfnamefont {D.}~\bibnamefont
  {van~der Marel}}, \bibinfo {author} {\bibfnamefont {J.}~\bibnamefont {van
  Mechelen}}, \ and\ \bibinfo {author} {\bibfnamefont {I.}~\bibnamefont
  {Mazin}},\ }\href {\doibase 10.1103/PhysRevB.84.205111} {\bibfield  {journal}
  {\bibinfo  {journal} {Phys. Rev. B}\ }\textbf {\bibinfo {volume} {84}},\
  \bibinfo {pages} {205111} (\bibinfo {year} {2011})}\BibitemShut {NoStop}%
\bibitem [{\citenamefont {Caviglia}\ \emph {et~al.}(2010)\citenamefont
  {Caviglia}, \citenamefont {Gabay}, \citenamefont {Gariglio}, \citenamefont
  {Reyren}, \citenamefont {Cancellieri},\ and\ \citenamefont
  {Triscone}}]{cavigliarashba}%
  \BibitemOpen
  \bibfield  {author} {\bibinfo {author} {\bibfnamefont {A.}~\bibnamefont
  {Caviglia}}, \bibinfo {author} {\bibfnamefont {M.}~\bibnamefont {Gabay}},
  \bibinfo {author} {\bibfnamefont {S.}~\bibnamefont {Gariglio}}, \bibinfo
  {author} {\bibfnamefont {N.}~\bibnamefont {Reyren}}, \bibinfo {author}
  {\bibfnamefont {C.}~\bibnamefont {Cancellieri}}, \ and\ \bibinfo {author}
  {\bibfnamefont {J.-M.}\ \bibnamefont {Triscone}},\ }\href@noop {} {\bibfield
  {journal} {\bibinfo  {journal} {Physical review letters}\ }\textbf {\bibinfo
  {volume} {104}},\ \bibinfo {pages} {126803} (\bibinfo {year}
  {2010})}\BibitemShut {NoStop}%
\bibitem [{\citenamefont {Spinelli}\ \emph {et~al.}(2010)\citenamefont
  {Spinelli}, \citenamefont {Torija}, \citenamefont {Liu}, \citenamefont
  {Jan},\ and\ \citenamefont {Leighton}}]{spinelli}%
  \BibitemOpen
  \bibfield  {author} {\bibinfo {author} {\bibfnamefont {A.}~\bibnamefont
  {Spinelli}}, \bibinfo {author} {\bibfnamefont {M.}~\bibnamefont {Torija}},
  \bibinfo {author} {\bibfnamefont {C.}~\bibnamefont {Liu}}, \bibinfo {author}
  {\bibfnamefont {C.}~\bibnamefont {Jan}}, \ and\ \bibinfo {author}
  {\bibfnamefont {C.}~\bibnamefont {Leighton}},\ }\href@noop {} {\bibfield
  {journal} {\bibinfo  {journal} {Physical Review B}\ }\textbf {\bibinfo
  {volume} {81}},\ \bibinfo {pages} {155110} (\bibinfo {year}
  {2010})}\BibitemShut {NoStop}%
\end{thebibliography}%

\end{document}